%% file: main.tex
\documentclass[sigconf, nonacm]{acmart}

\usepackage{algorithm}
\usepackage{listings}
\usepackage{hyperref}
\usepackage[nameinlink]{cleveref}
\usepackage{subfig}
\usepackage{colortbl}
\usepackage{xcolor}
\usepackage{pifont}
\usepackage{booktabs}
\usepackage{multirow}
\usepackage{multicol}
\usepackage{makecell}
\usepackage{float}
\usepackage{dblfloatfix}
\usepackage{soul} 
\colorlet{lightgrey}{gray!20} 
\sethlcolor{lightgrey}
\usepackage{titlesec} 
\usepackage{tikz}
\usepackage{multicol} 
\usepackage{enumitem}

\newcommand*\rev[1]{\textcolor{black}{#1}}

\newenvironment{revision}{%
}{%
}%

\newcommand{\changedimage}[1]{%
  #1%
}

\newcommand*\circled[1]{\tikz[baseline=(char.base),font=\bfseries\sffamily]{
            \node[shape=circle,draw,inner sep=0.5pt] (char) {#1};}}

\titlespacing*{\paragraph}{0pt}{0pt}{1ex}
\titlespacing*{\section}{0pt}{0.7\baselineskip}{0.5\baselineskip}
\titlespacing*{\subsection}{0pt}{0.5\baselineskip}{0.2\baselineskip}

\titleformat{\paragraph}[runin]
  {\normalfont\normalsize\bfseries}  
  {}                                 
  {0pt}                              
  {}                                 

\newcommand\vldbdoi{XX.XX/XXX.XX}
\newcommand\vldbpages{XXX-XXX}
\newcommand\vldbvolume{14}
\newcommand\vldbissue{1}
\newcommand\vldbyear{2020}
\newcommand\vldbauthors{\authors}
\newcommand\shorttitle{PystachIO: Efficient Distributed GPU Query Processing with PyTorch over Fast Networks \& Fast Storage}
\newcommand\vldbtitle{\shorttitle}
\newcommand\vldbavailabilityurl{https://osf.io/yv4na/overview?view_only=5e2d65473b9248bfbb86946e9e6fd44a}
\newcommand\vldbpagestyle{plain}

\begin{document}

\title[PystachIO: Efficient Distributed GPU Query Processing with PyTorch over Fast Networks \& Fast Storage]{PystachIO: Efficient Distributed GPU Query Processing \\with PyTorch over Fast Networks \& Fast Storage}
\subtitle{[Flavor: Systems]}

\author{Jigao Luo}
\orcid{0009-0005-2263-1959}
\affiliation{%
  \institution{TU Darmstadt}
  \country{}
}
\authornote{Equal contribution.}

\author{Nils Boeschen}
\orcid{0000-0001-8654-5738}
\affiliation{
  \institution{TU Darmstadt \& hessian.AI}
  \country{}
}
\authornotemark[1]

\author{Muhammad El-Hindi}
\orcid{0000-0002-1825-0097}
\affiliation{%
  \institution{TU Munich}
  \country{}
}

\author{Carsten Binnig}
\orcid{0000-0002-2744-7836}
\affiliation{
  \institution{TU Darmstadt \& hessian.AI \& DFKI Darmstadt}
  \country{}
}

\input{section/00_abstract}

\maketitle

\pagestyle{\vldbpagestyle}
\begingroup\small\noindent\raggedright\textbf{PVLDB Reference Format:}\\
\vldbauthors. \vldbtitle. PVLDB, \vldbvolume(\vldbissue): \vldbpages, \vldbyear.\\
\href{https://doi.org/\vldbdoi}{doi:\vldbdoi}
\endgroup
\begingroup
\renewcommand\thefootnote{}\footnote{\noindent
This work is licensed under the Creative Commons BY-NC-ND 4.0 International License. Visit \url{https://creativecommons.org/licenses/by-nc-nd/4.0/} to view a copy of this license. For any use beyond those covered by this license, obtain permission by emailing \href{mailto:info@vldb.org}{info@vldb.org}. Copyright is held by the owner/author(s). Publication rights licensed to the VLDB Endowment. \\
\raggedright Proceedings of the VLDB Endowment, Vol. \vldbvolume, No. \vldbissue\ %
ISSN 2150-8097. \\
\href{https://doi.org/\vldbdoi}{doi:\vldbdoi} \\
}\addtocounter{footnote}{-1}\endgroup

\ifdefempty{\vldbavailabilityurl}{}{
\vspace{.3cm}
\begingroup\small\noindent\raggedright\textbf{PVLDB Artifact Availability:}\\
The source code, data, and/or other artifacts have been made available at \url{\vldbavailabilityurl}.
\endgroup
}

\input{section/01_introduction}

\input{section/03_overview}

\input{section/05_network}
\input{section/07_storage}

\input{section/10_network_and_storage}

\input{section/60_evaluation}


\input{section/70_related}
\input{section/80_conclusion}

\balance



\bibliographystyle{ACM-Reference-Format}
\bibliography{the_bibliography}

\newpage

\end{document}

%% file: section/00_abstract.tex

\begin{abstract}
The AI hardware boom has led modern data centers to adopt HPC-style architectures centered on distributed, GPU-centric computation.
Large GPU clusters interconnected by fast RDMA networks and backed by high-bandwidth NVMe storage enable scalable computation and rapid access to storage-resident data.
Tensor computation runtimes (TCRs), such as PyTorch, originally designed for AI workloads, have recently been shown to accelerate analytical workloads.
However, prior work has primarily considered settings where the data fits in aggregated GPU memory.
In this paper, we systematically study how TCRs can support scalable, distributed query processing for large-scale, storage-resident OLAP workloads.
Although TCRs provide abstractions for network and storage I/O, naive use often underutilizes GPU and I/O bandwidth due to insufficient overlap between computation and data movement.
As a core contribution, we present PystachIO, a \rev{prototype of a} PyTorch-based distributed OLAP engine that combines fast network and storage I/O with key optimizations to maximize GPU, network, and storage utilization.
Our evaluation shows up to 3$\times$ end-to-end speedups over existing distributed GPU-based query processing approaches.
\end{abstract}

%% file: section/01_introduction.tex

\section{Introduction} \label{sec:introduction}

\begin{figure}[t]
    \centering
    \includegraphics[width=0.9\columnwidth]{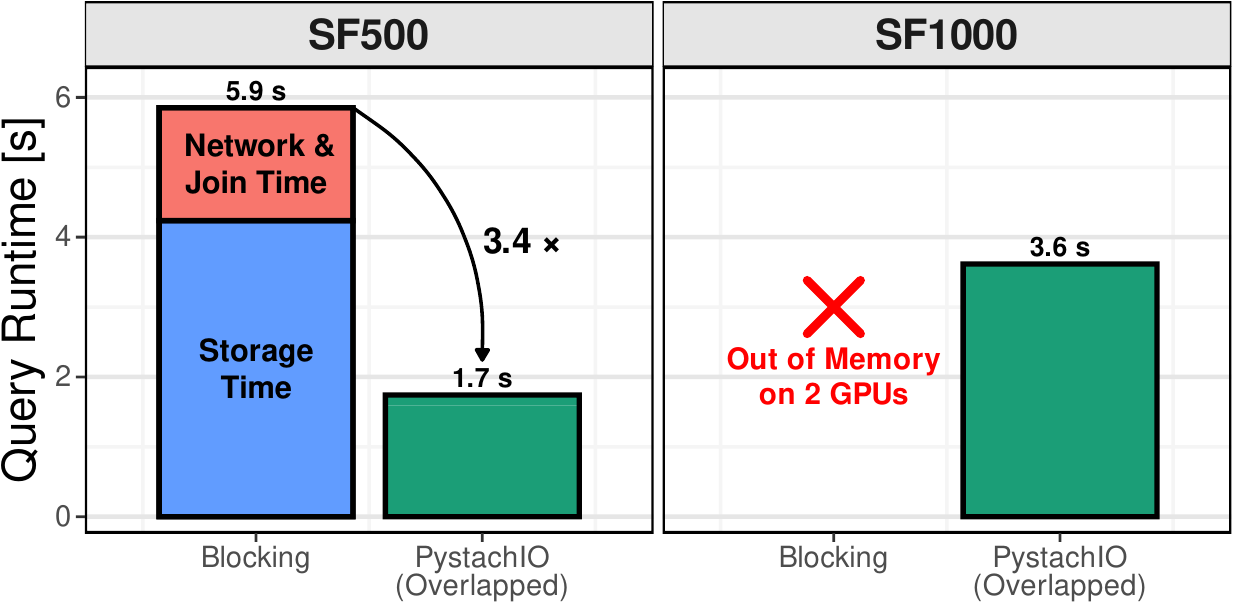}
    \caption{TPC-H Q3 runtime at scale factors (SF) 500 and 1000 on two GPUs over SSD-resident, non-co-partitioned tables. A blocking PyTorch-style baseline executes sequentially, leading to long runtimes and out-of-memory errors. In contrast, PystachIO overlaps storage I/O, networking, and computation to reduce query time by over 3$\times$.}
    \label{fig:fig1}
    \vspace{-5ex}
    \Description{} 
\end{figure}

\paragraph{The Rise of AI Hardware.}  
The recent AI boom has driven massive investments in new data centers built for large-scale, distributed AI training~\cite{mckinsey2025,Amazon_2025}.
These facilities increasingly adopt HPC-style architectures built around high-performance GPU clusters interconnected by fast, low-latency networks and backed by high-bandwidth NVMe storage~\cite{DBLP:journals/csur/YeGHSWLZW24,DBLP:conf/itnac/Gupta24}.
As GPUs have become the primary compute units for training and inference, hyperscalers increasingly rely on their high compute density and acceleration capabilities.
Consequently, modern data center infrastructure enables efficient GPU-centric distributed execution and fast access to massive storage-resident datasets.
Beyond AI workloads, these investments are accelerating the commoditization of large-scale GPU computing, making high-performance GPU infrastructure broadly available and opening new opportunities for other data-intensive workloads, including analytical query processing.

\paragraph{Leveraging AI Abstractions for Query Processing.}  
The rapid adoption of large-scale GPU computing is also fueled by simple, expressive tensor computation runtimes (TCRs) such as PyTorch~\cite{pytorch_website}.
Although these abstractions were initially designed for developing and training AI models, recent work showed that they can also be used to implement analytical database systems~\cite{DBLP:journals/pvldb/HeNBSSPCCKI22}.
By mapping relational operators (e.g., filters, joins, and aggregations) to the vectorized tensor operations provided by PyTorch, these systems enable high-performance processing for OLAP workloads on GPUs.

\paragraph{Limitations of Existing Tensor-Based Engines.}  
Existing TCR-based systems for analytical processing primarily target single-node setups and assume that table data fits entirely in GPU memory.
However, modern OLAP workloads often exceed the memory capacity of a single GPU, making such single-node in-memory designs cost-prohibitive and impractical.
Recent work has begun to explore distributed query processing with TCRs to handle larger datasets~\cite{DBLP:journals/corr/abs-2506-09226}, but it still assumes that the workload fits into the aggregate memory of the GPU cluster.
Consequently, scaling to larger workloads requires provisioning additional accelerator nodes, which increases costs and operational complexity.

\paragraph{Distributed, Out-Of-Memory OLAP With TCRs.}  
To address these limitations and enable large-scale, storage-resident OLAP on GPUs, we investigate how tensor computation runtimes can support distributed, out-of-memory query processing.
In contrast to prior work, our approach leverages the full AI data center stack:
fast RDMA-capable networks for distributed computation and high-bandwidth NVMe SSDs for out-of-memory execution.
A key challenge is that naively reusing existing TCR abstractions for network and storage I/O, as is typical in large-scale model training, yields a blocking execution model in which data is first loaded in bulk before processing begins.
While this pattern is suitable for compute-intensive, iterative training workloads, it results in low query performance and even out-of-memory (OOM) errors for I/O-intensive OLAP workloads, as shown by the left bars in \Cref{fig:fig1}.

\paragraph{PystachIO: Efficient Large-Scale OLAP With TCRs.}
We present \emph{PystachIO}, a \rev{prototypical} GPU-accelerated distributed query engine that enables fast out-of-memory OLAP with tensor computation runtimes, as shown by the right bars in \Cref{fig:fig1}.
PystachIO is built on PyTorch and is designed to efficiently leverage high-performance networks and NVMe storage.
Our key observation is that achieving high performance requires careful utilization of network and storage bandwidth to avoid bottlenecks and idle GPU cycles.
PystachIO addresses this by overlapping computation with storage and network I/O, while reducing synchronization overheads and optimizing memory usage for large datasets.

\paragraph{Contributions and Outline.}  
In summary, this paper makes the following contributions:  
(1) We analyze why naive use of TCR I/O abstractions fails to fully exploit the capabilities of fast networks and storage (\Cref{sec:background}).  
(2) We introduce optimizations that enable query processing to approach the utilization limits of network and storage bandwidth by effectively overlapping I/O with computation (\Cref{sec:network,sec:storage}).  
(3) We realize these techniques in PystachIO, a distributed query engine that carefully coordinates storage and networking to maximize end-to-end performance (\Cref{sec:network_and_storage}). 
(4) We evaluate PystachIO on analytical workloads, demonstrating performance close to the theoretical hardware bounds (\Cref{sec:evaluation}).

We conclude by discussing related work in \Cref{sec:related} and summarizing the main insights in \Cref{sec:conclusion}.

%% file: section/03_overview.tex

\section{System Overview \& Challenges} \label{sec:background}
In this section, we provide an overview of PystachIO's system architecture and discuss the challenges of leveraging TCRs to build an efficient and scalable distributed database system.

\subsection{System Architecture} \label{sec:arch}
\begin{figure}[t]
\includegraphics[width=.9\columnwidth]{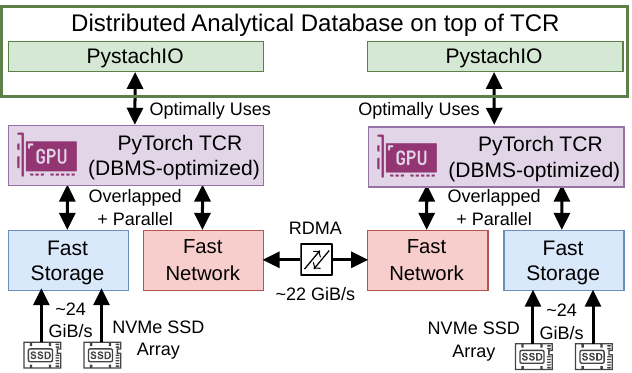}
\caption{PystachIO: a database architecture that exploits fast network and storage hardware in AI data centers on top of the PyTorch TCR. Naive use of the TCR does not achieve optimal performance. PystachIO adds database-specific optimizations, including overlapping compute and I/O and parallel storage access, to enable highly efficient query processing.}
\label{fig:pystachio_arch}
\Description{}
\vspace{-4ex}
\end{figure}

\citeauthor{DBLP:journals/pvldb/HeNBSSPCCKI22}~\cite{DBLP:journals/pvldb/HeNBSSPCCKI22} pioneered the idea of using AI-oriented TCRs for GPU-accelerated query processing. 
PystachIO advances this direction by implementing a distributed TCR-based engine that exploits the fast networks and storage technologies commonly used in modern AI data centers: RDMA-enabled networking and high-bandwidth NVMe storage.
\Cref{fig:pystachio_arch} illustrates the overall system design.
\rev{Like prior work, PystachIO employs a column-oriented storage layout in which each column is represented as a tensor, to enable the use of optimized parallel TCR primitives.
The system further partitions these tensors horizontally into chunks, which enables overlapping I/O and computation and is key to PystachIO's efficient use of fast storage and networking.}

\paragraph{Fast Networking and Storage Using TCRs.}
TCRs expose abstractions for high-speed networking and storage, originally designed for distributed AI workloads.
However, these abstractions do not deliver high performance for database workloads out-of-the-box, since data movement rather than computation dominates database execution.
Below, we outline the relevant TCR primitives for fast networking and storage and highlight the challenges that arise when using them for database workloads.
We then summarize how PystachIO addresses these challenges through careful use and internal extensions of TCRs.
With these improvements, TCRs such as PyTorch can serve as a high-performance substrate for distributed database systems without requiring the design of systems based on low-level, system-specific interfaces.

\subsection{Fast Networking With TCRs}
\begin{figure}
\includegraphics[width=0.85\columnwidth]{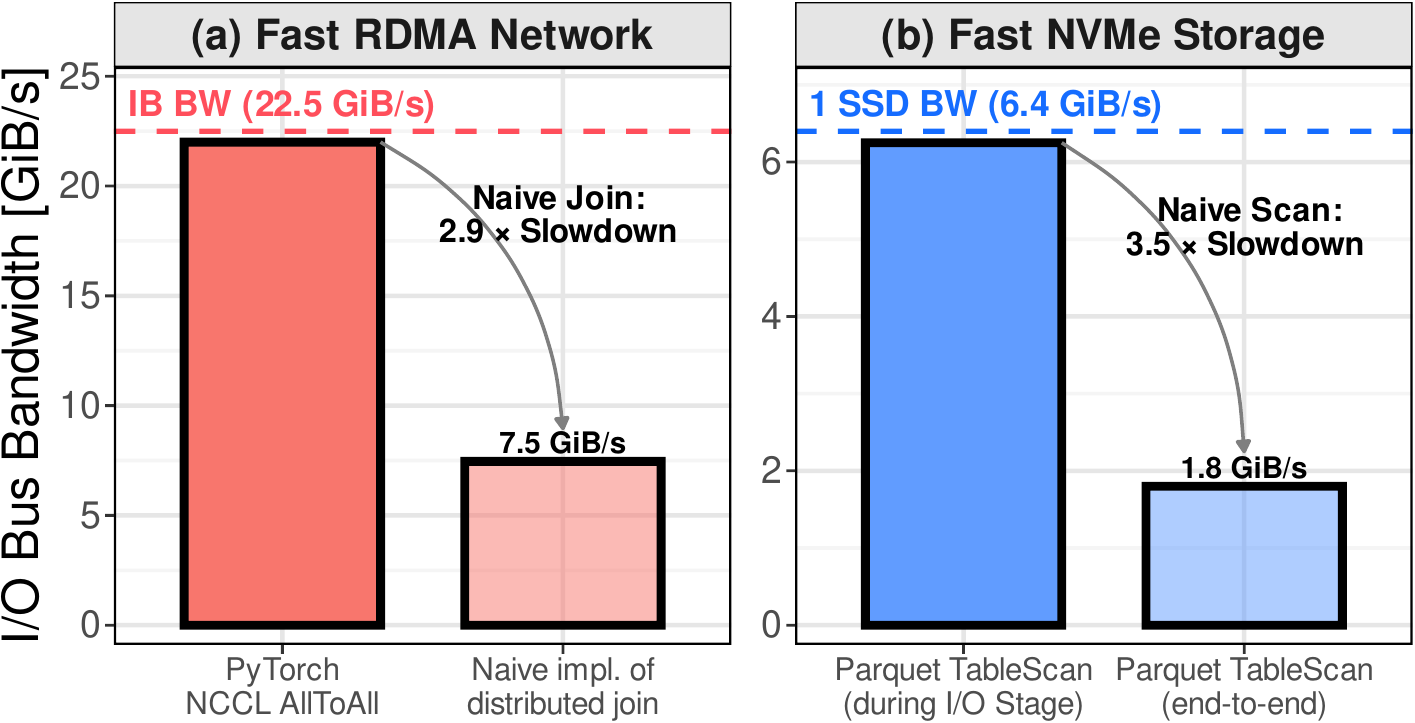}
\caption{TCRs provide fast networking and storage primitives:
(a) Performance of PyTorch's \texttt{AllToAll} collective versus a naive distributed join.
(b) SSD read performance with GDS enabled, measured during the I/O stage of a scan versus the end-to-end full scan.}
\label{fig:basic_net_and_storage}
\Description{}
\vspace{-5ex}
\end{figure}

PyTorch provides efficient networking capabilities typically used during distributed AI training.
Specifically, its distributed communications layer offers collective and point-to-point primitives, backed by the NVIDIA Collective Communications Library (NCCL)~\cite{nccl_website}, which enables direct network I/O between GPU memory and RDMA-capable network hardware via GPUDirect RDMA~\cite{gdr_website}.
As shown in \Cref{fig:basic_net_and_storage}\textup{(a)}, NCCL can saturate high-speed networks and reach near-peak bandwidth when invoked through PyTorch.

\paragraph{Challenges in TCR-Based Query Processing.}
However, naively applying collectives to implement a distributed join results in suboptimal bandwidth (see \Cref{fig:basic_net_and_storage}\textup{(a)} second bar) and uneven network utilization.
The primary issue is that communication is blocking, leaving GPUs idle during data shuffling.
While blocking communication is acceptable in AI workloads, where computation dominates the overall cost, it becomes a bottleneck for database workloads, where network transfer time is comparable to processing time.

\paragraph{Network I/O in PystachIO.}
PystachIO builds on PyTorch's distributed communication layer with the NCCL backend, to shuffle data efficiently for distributed processing across GPUs over RDMA networks.
Instead of relying on PyTorch's default blocking operations, we introduce new scheduling techniques that overlap communication with computation, effectively hiding computation latency.
Additionally, PystachIO optimizes runtime memory allocation and eliminates unnecessary synchronization, sustaining a consistently high network bandwidth during query execution.
We discuss the design of PystachIO's network layer in \Cref{sec:network}.

\subsection{Fast Storage With TCRs}

PyTorch-based TCRs typically access storage-resident tabular data via libraries such as \texttt{pandas}~\cite{pandas_website}, \texttt{NumPy}~\cite{numpy_website}, and RAPIDS cuDF~\cite{libcudf}, supporting file formats including CSV and Apache Parquet~\cite{10.5555/3012315}.
CPU-based readers like \texttt{pandas} and \texttt{NumPy} perform parsing and data movement on the CPU and thus cannot saturate NVMe bandwidth~\cite{DBLP:journals/pacmmod/KuschewskiSAL23,DBLP:journals/corr/abs-2504-15247}.
In contrast, cuDF supports GPUDirect Storage (GDS) \cite{gds_website,kvikio_website}, which enables direct I/O between NVMe SSDs and GPU memory, followed by GPU kernels for tasks such as Parquet decompression.
As \Cref{fig:basic_net_and_storage}\textup{(b)} shows, a GDS-enabled cuDF table scan over a Parquet file can saturate a single SSD during the I/O stage.

\paragraph{Challenges in TCR-Based Query Processing.}
However, as with networking, PyTorch- and cuDF-based applications typically perform storage I/O in a blocking fashion.
As a result, kernels that process data after I/O, such as result filtering, run only after the full I/O phase has completed, leaving the SSD idle during GPU processing.
This serialized execution reduces average storage bandwidth by roughly $3.5\times$, as shown in \Cref{fig:basic_net_and_storage}\textup{(b)} (right bar).

\paragraph{Storage I/O in PystachIO.}
To enable efficient storage access in PystachIO, we redesign the GPU-based table scan by replacing blocking file-at-a-time I/O with a fully overlapped pipeline that interleaves storage I/O and GPU computation.
Further optimizations include metadata caching to eliminate repeated CPU-side overheads and pinned-memory-based synchronization that removes hidden GPU stalls.
These techniques transform the table scan into a fully pipelined process that saturates SSD bandwidth and keeps the GPU consistently busy, enabling efficient storage utilization for TCR-based query processing.
We discuss the design of PystachIO's storage layer in Section \ref{sec:storage}.

%% file: section/05_network.tex

\section{Distributed Execution With TCRs} \label{sec:network}
Traditionally, only specialized, hand-optimized DBMSs have been able to fully exploit high-bandwidth, low-latency networks, such as RDMA-capable interconnects in AI data centers (e.g., \cite{DBLP:journals/corr/abs-2508-05029,DBLP:conf/adms/GaoS21,DBLP:journals/pvldb/BarthelsAHSM17}).
As discussed in the previous section, TCRs such as PyTorch provide high-level abstractions over networking primitives, making it easier for developers to leverage and saturate these fast networks.
However, naive use of these networking abstractions in query processing underutilizes the underlying hardware.
In this section, we present optimizations that enable TCR-based database systems to maximize network bandwidth by overlapping computation and communication.
We then use these techniques to design highly efficient distributed hash joins, one of the most expensive operations in distributed query processing.

\subsection{Naive Approach to Distributed Joins}

\begin{figure}
\includegraphics[width=\columnwidth]{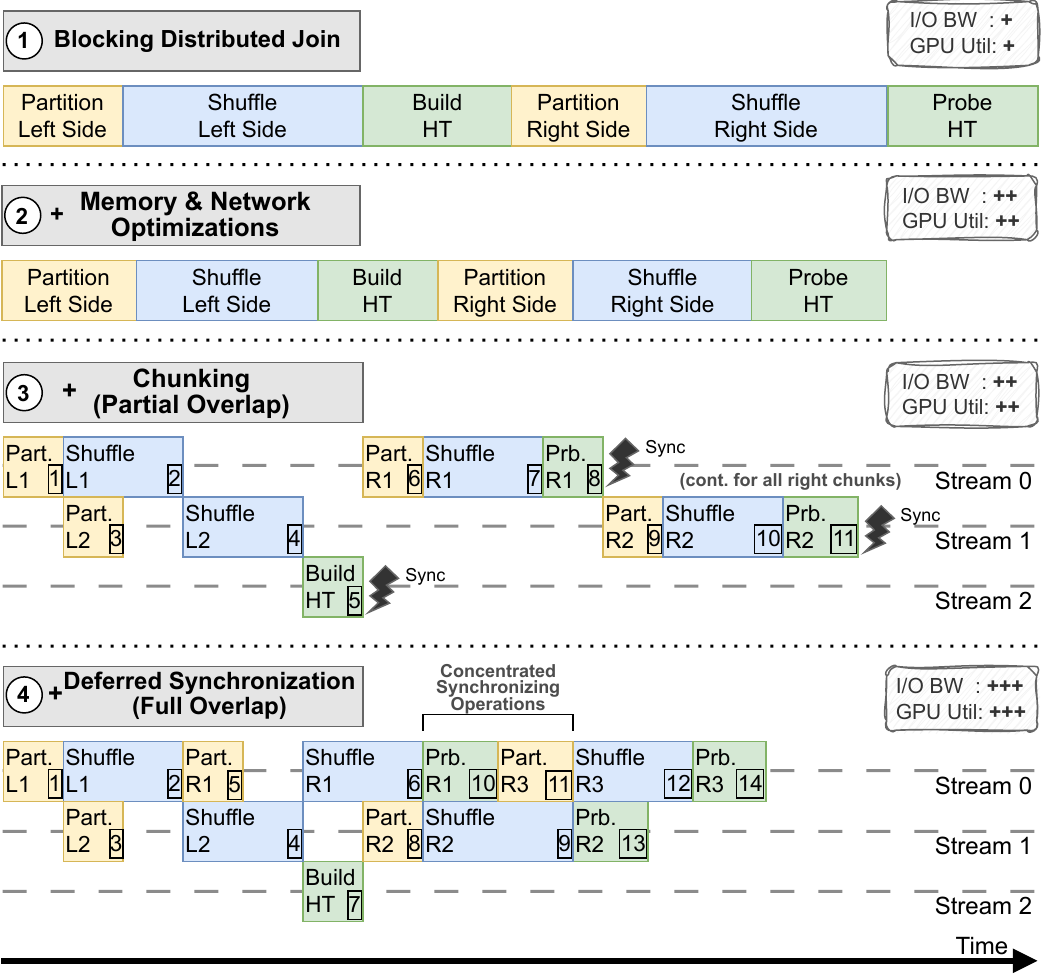}
\caption{
Naive and optimized variants of a distributed hash join in PyTorch.
\rev{The blocking variant is the most natural mapping to PyTorch, but also the least efficient.
Later variants incorporate the overlap of phases and can significantly improve runtime.}
For each variant, the expected GPU and network I/O utilization is shown in the top-right inset, and the scheduling order of operations (for the overlapped variants) is annotated in the bottom-right corner.} 
\label{fig:naive_vs_overlap_figure}
\vspace{-4ex}
\Description{TODO}
\end{figure}

In this section, we demonstrate how PyTorch interfaces can be leveraged to implement a distributed hash join.
For the network I/O between GPUs over RDMA-capable networks, we employ NCCL.
We use the \emph{AllToAll} collective, which is implemented as a wrapper around the respective point-to-point transfers.

\paragraph{Anatomy of a Distributed Hash Join.}
A distributed hash join over two tables can be conceptually decomposed into six steps (see top of \autoref{fig:naive_vs_overlap_figure}).
The local left table is first partitioned according to the join key, after which the resulting partitions are shuffled across the network.
Each node then builds a hash table from the left-hand partitions it receives.
The local right table undergoes the same procedure: partitioning and shuffling based on the join key.
Finally, each node probes the hash table using the right-hand partitions and materializes or passes on the resulting join tuples.

\paragraph{A Naive Implementation in PyTorch.}
Using PyTorch's tensor operations and network primitives, a straightforward blocking variant, as shown as \emph{Blocking Distributed Join} \circled{1} in \autoref{fig:naive_vs_overlap_figure}, is simple and natural to implement.
Partitioning can be performed through vectorized modulo operations combined with masked selection.
To communicate partition sizes across nodes, an \emph{AllToAll} collective can be used, preparing the receivers for incoming tensor data.
The actual shuffling of columns can be implemented via NCCL's \emph{AllToAll} collective for each partitioned column, assuming that communication efficiently utilizes direct GPU-to-NIC I/O and fully leverages the underlying high-speed interconnect.
Hash table construction and probing can be mapped to scatter and gather operations, as proposed in~\cite{DBLP:journals/pvldb/HeNBSSPCCKI22}.

\begin{figure}
\subfloat[Runtime of join variants.]{
    \includegraphics[width=\columnwidth, trim={0cm 0.9cm 0 0.9cm},clip]{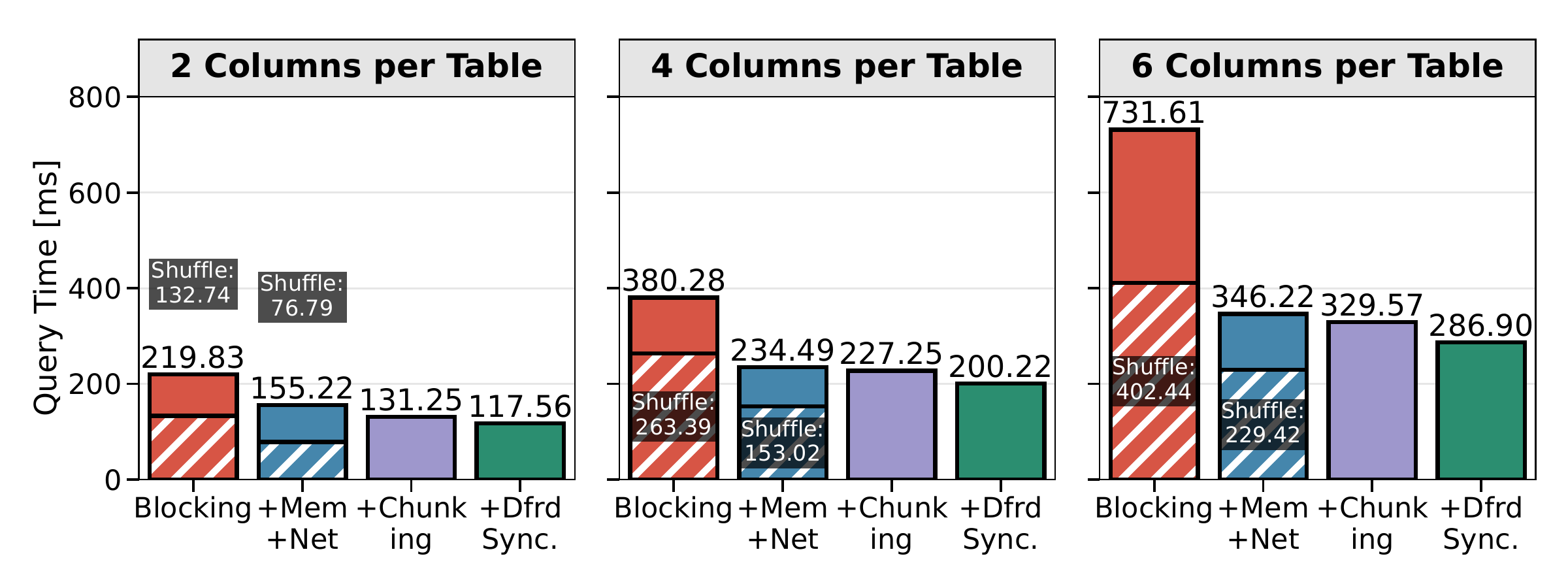}
}\\\vspace{-1em}
\subfloat[Network BW (total data sent/received per node divided by runtime).]{
    \includegraphics[width=\columnwidth, trim={0cm 0.9cm 0 0.9cm},clip]{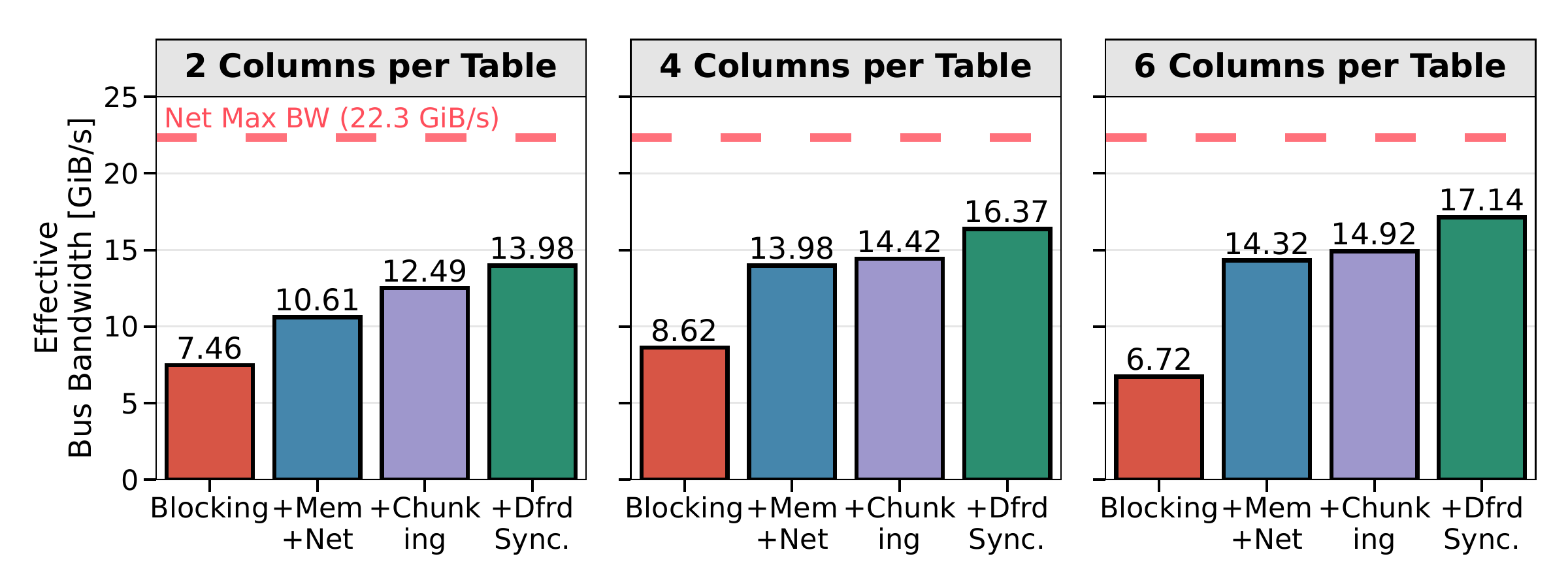}
}\\\vspace{0.5em}
\caption{Distributed join performance on 2 GPU nodes for varying numbers of 8-byte columns (120M/320M build/probe tuples, hit ratio 0.5). Hatched bars show network shuffle time for the blocking variants.}
\label{fig:shuffle_scalability2}
\vspace{-3ex}
\Description{TODO}
\end{figure}

\paragraph{Analysis of the Blocking Join.}
We evaluate the performance of the blocking variant by shuffling tables with varying widths over a 200G RoCE network\footnote{All details on the hardware setup are provided in \Cref{sec:evaluation}.}.
We report the effective I/O-bus bandwidth, defined as the total data received (or sent) per node divided by the end-to-end join runtime.
The experiment shows that the naive implementation, labeled \emph{Blocking} (red) in \Cref{fig:shuffle_scalability2}, sustains less than 50\% of the maximum network bandwidth.
This suboptimal performance arises from two main inefficiencies.

\paragraph{(1) Low Utilization of Available Network Bandwidth.}
Because the networking backends of TCRs are designed to balance resource usage and performance (e.g., NCCL launches only a limited number of GPU thread blocks by default), parameter tuning is required to approach the bandwidth limits of high-speed networks.
In our NCCL-based setup, we tune both the per-message size (512~KiB) and the number of GPU-executed blocks (32 NCCL channels).
This tuning enables us to achieve near-peak network utilization, as observed in the \emph{AllToAll} results in \Cref{fig:basic_net_and_storage}.

\paragraph{(2) Runtime Memory Allocation in PyTorch.}
PyTorch allocates memory dynamically by default.
While this behavior is well-suited for compute-bound AI training, database operators are often less compute-intensive and therefore more sensitive to memory allocation overhead.
Thus, we enable allocating dynamically sized, data-dependent tensors by configuring PyTorch to use the RAPIDS Memory Manager (RMM) allocator~\cite{rmm}.
RMM allocates a large block of device memory once and then serves smaller allocations from that pool with low overhead.

\paragraph{Performance Impact of Low-Level Optimizations.}
The low-level optimizations, tuning network parameters, and improving memory allocation can be applied without modifying the join algorithm itself, as shown in \Cref{fig:naive_vs_overlap_figure} \circled{2}.
They already reduce the runtime of the naive blocking join by 1.5--2$\times$, as indicated by \emph{+Mem+Net} (blue) in \Cref{fig:shuffle_scalability2}.
Because they require no algorithmic changes, they are an effective first step in optimizing distributed joins.

\begin{figure*}
    \includegraphics[trim={0 0.5em 0 0},clip,width=.3\columnwidth]{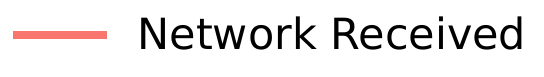}\\\vspace{-0.4em}
    \includegraphics[height=2.84cm]{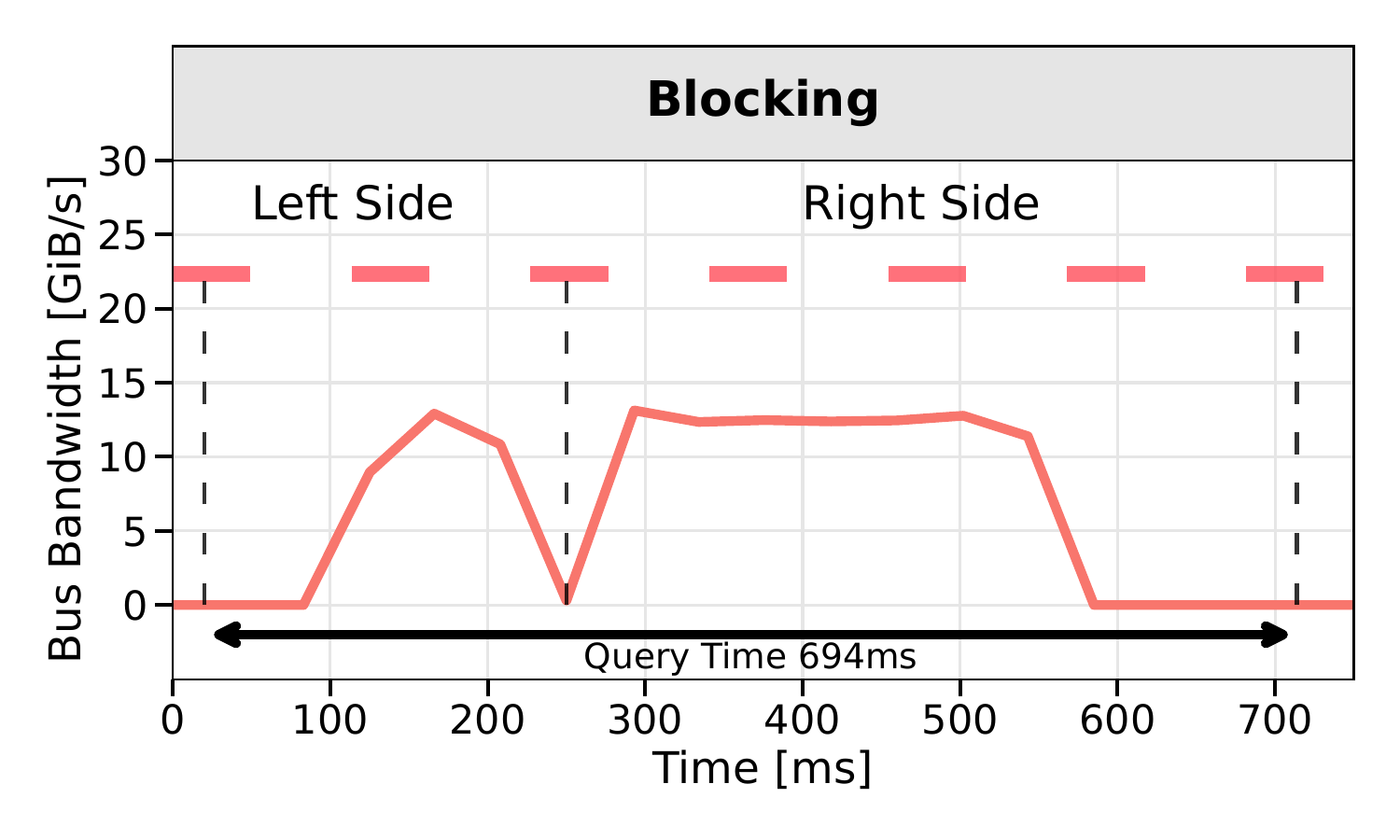}\hfill
    \includegraphics[trim={6.5em 0 0 0},clip,height=2.84cm]{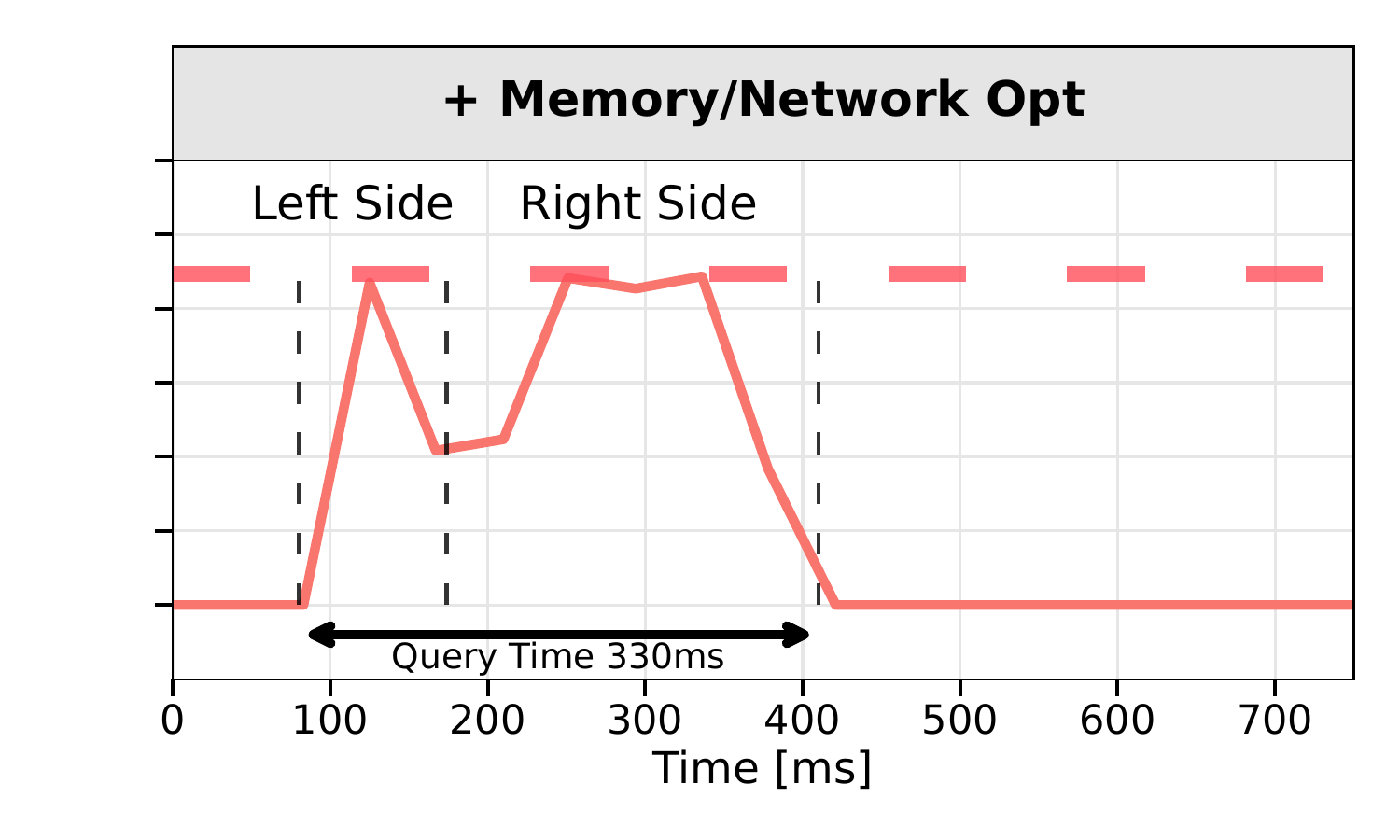}\hfill
    \includegraphics[trim={6.5em 0 0 0},clip,height=2.84cm]{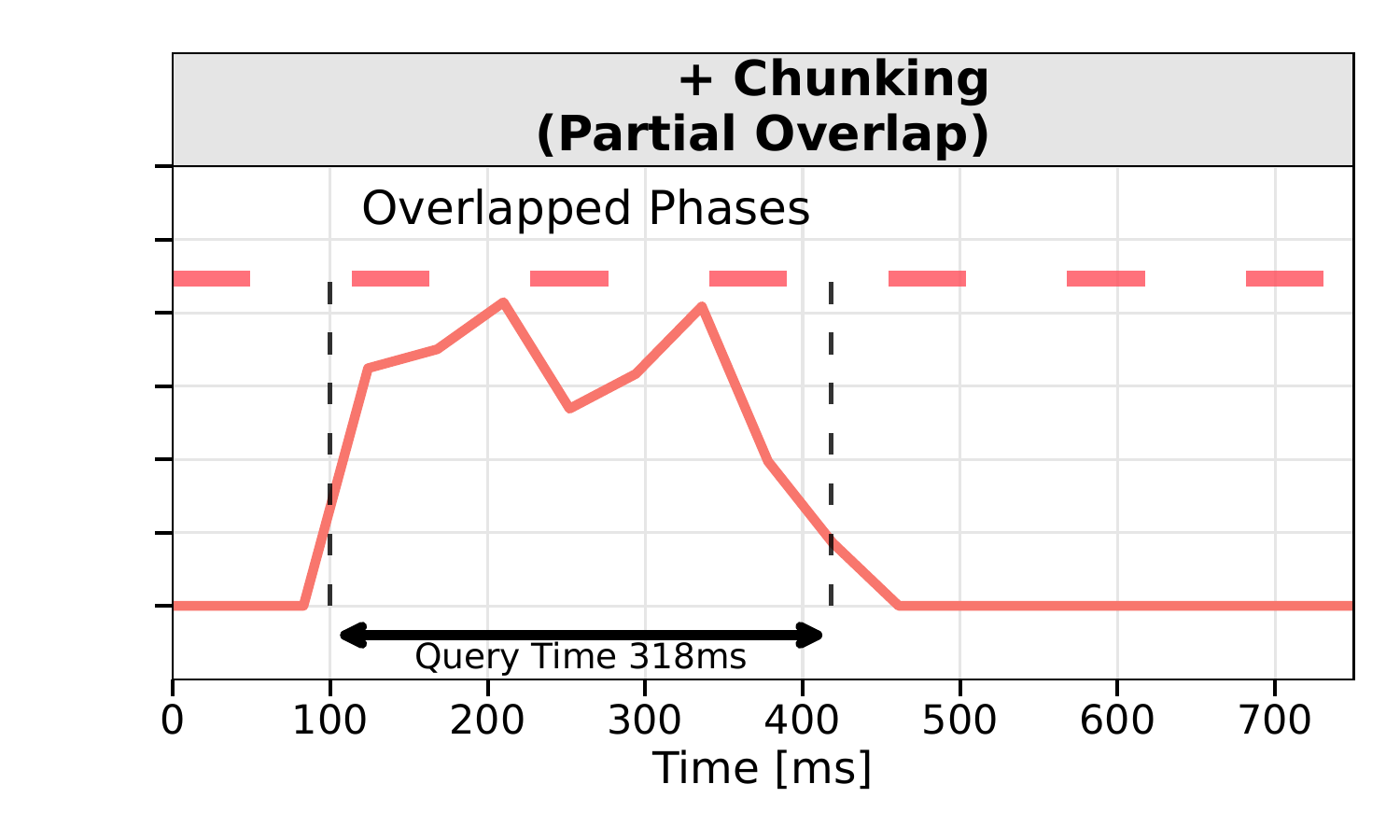}\hfill
    \includegraphics[trim={6.5em 0 0 0},clip,height=2.84cm]{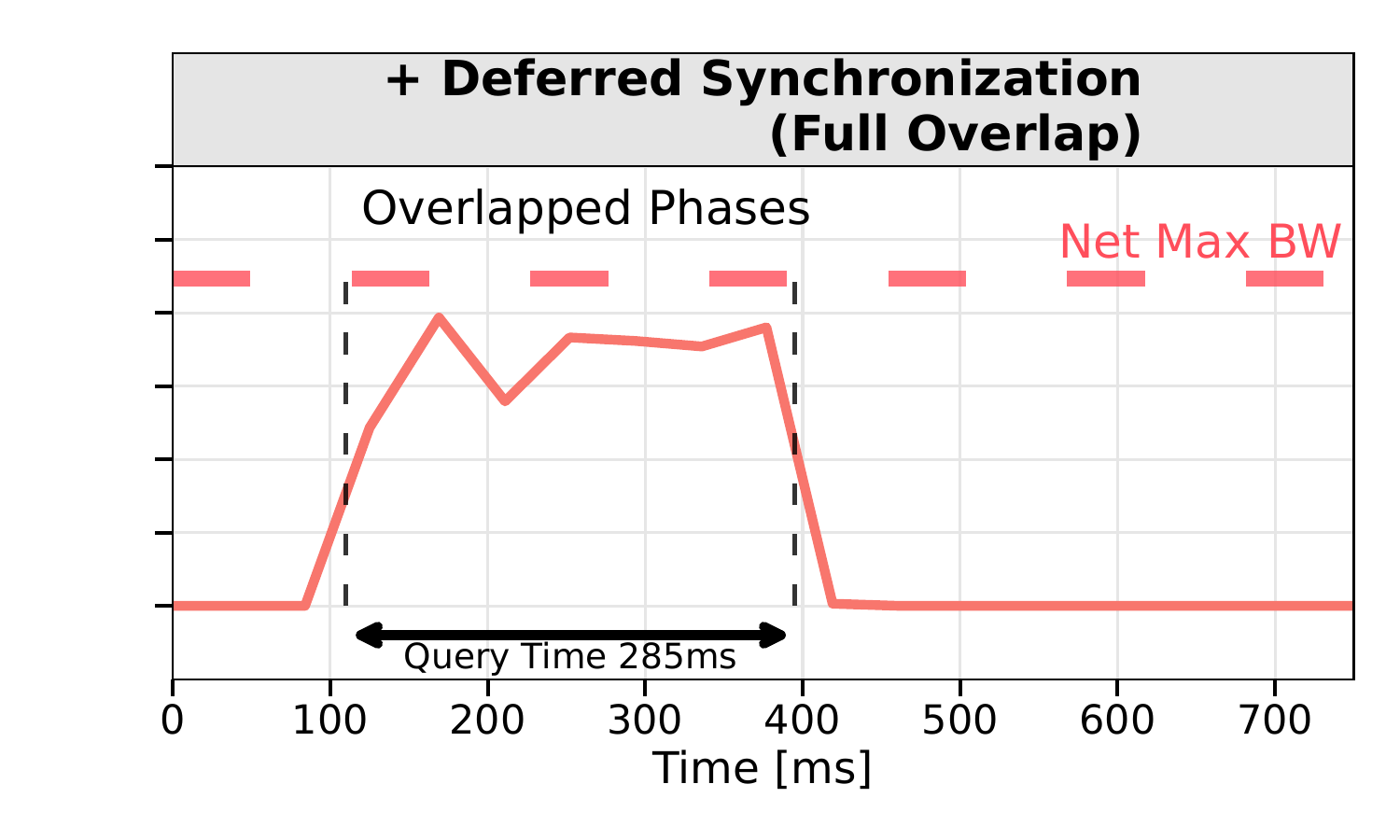}
\caption{Timeline of bandwidth monitoring of one involved NIC during different distributed join variants. 2 Nodes, 120M/160M build/probe side tuples in total, 6 Columns, join hit ratio 0.5.}
\label{fig:beemon_shuffle}
\vspace{-3ex}
\Description{TODO}
\end{figure*}

\subsection{Optimization: Chunking} 
\label{sec:overlap_joins}

As shown earlier, a naive blocking approach for distributed joins underutilizes the bandwidth of modern networks.
Hence, we investigate how to implement a non-blocking distributed join in PyTorch.
Instead of blocking, we overlap communication and computation to improve network utilization and reduce overall query runtime.

\paragraph{Partial Overlap Using Chunking.}
The compute and memory resources of modern datacenter GPUs enable concurrent execution of network communication and local computation.
We leverage this capability to design an overlapped variant of the distributed join that improves end-to-end runtime.
To overlap I/O and computation, all phases of the distributed join must operate on data chunks rather than PyTorch's default full-column tensors.
Chunking enables, for example, the partitioning of one chunk to run concurrently with the network shuffling of another.
When implemented effectively, such an overlap significantly reduces idle periods and shortens overall processing time, as we show later.

\paragraph{Scheduling Overlap With GPU Streams.}
To overlap different phases of the join, we leverage the ability of modern GPUs to schedule operations across multiple execution queues (CUDA streams).
These streams enable the asynchronous launch of compute kernels and network primitives, allowing operations issued to different streams to run concurrently, purely
from GPU internal overlap, without any CPU thread parallelism.
As illustrated by \emph{Chunking} \circled{3} in \autoref{fig:naive_vs_overlap_figure}, we use standard data-parallel execution across streams: for each data chunk, partitioning, shuffling, and hash table operations are scheduled sequentially within one stream, while other chunks are assigned to separate streams.
For example, \Cref{fig:naive_vs_overlap_figure} shows how operations on two chunks of the left-side table can overlap: the shuffling of the first left chunk \texttt{L1} and the partitioning of \texttt{L2} are executed concurrently. 
Because the hash table is built on a fully materialized table~\cite{DBLP:journals/pvldb/HeNBSSPCCKI22}, we build it once per node in a separate stream after all left-side chunks have been received.
The scheduling order for this variant (annotated as numbers in the lower right corner) follows the per-chunk pipeline: for chunk \texttt{R1}, partition, shuffle, and probe are scheduled back-to-back.
The performance impact of overlap via \emph{Chunking} over the optimized blocking approach is modest, as shown by the purple bars in \autoref{fig:shuffle_scalability2}.

\rev{However, since this approach enables \emph{streaming} received chunks to subsequent operators and dropping data after processing (keeping only the very necessary intermediate results in accelerator memory), chunking allows for relieving memory pressure drastically, and consequently to join large tables in fewer accelerators.}

\paragraph{Remaining Bandwidth Limitations.}
The achieved bandwidth still falls short of the network's capacity.
This observation is evident from the network utilization timelines in \autoref{fig:beemon_shuffle}, which show the activity of a single NIC during the shuffling process.
As expected, the \emph{Blocking} variants (first two timelines) exhibit pronounced gaps between processing the two input tables, during which network traffic temporarily drops.
The \emph{Chunking} variant eliminates this large gap and reduces overall runtime, yet still shows considerable bandwidth fluctuations, especially in the second half of execution, indicating inefficient network utilization.

\paragraph{Stream-Synchronizing Operations Limit Overlap.}
One root cause of the remaining bandwidth underutilization is reduced overlap between operations due to implicit stream synchronization.
Synchronizing a GPU stream (of kernels or memory transfers) blocks the CPU until all previously enqueued operations in that stream have completed.
Many TCR primitives introduce such synchronizations implicitly because they must return tensors with data-dependent metadata (e.g., size) that is only known after execution on the device.
A representative example is \texttt{masked\_select}, which selects tensor entries based on a boolean mask and is used for partitioning as well as for hash table probing~\cite{DBLP:journals/pvldb/HeNBSSPCCKI22}.
The size of its output tensor depends on the mask, but runtimes such as PyTorch require the returned tensor to be correctly sized when the call returns, even if the actual data is populated asynchronously.
To determine this size, the TCR synchronizes the corresponding stream and materializes the necessary metadata on the CPU, thereby blocking further kernel launches from the CPU thread.
\rev{Such synchronizations are inherent to distributed query processing, which involves operations with data-dependent result sizes (e.g., selections, joins, group-by operations). In the case of the distributed join, the partitioning and hash table
operations include stream-synchronizing primitives, while the NCCL \emph{AllToAll} collective itself can execute fully asynchronously (after metadata has been exchanged), as long as no synchronizing operation follows it in the same stream.}

These synchronizations are particularly harmful during hash table probing on received chunks.
If probing is scheduled per chunk directly after shuffling, the synchronizing probe step blocks until all earlier operations in that stream (partitioning and shuffling) have completed, as illustrated by the lightning symbol for the probe of \texttt{R1} in \Cref{fig:naive_vs_overlap_figure} (\emph{Chunking} \circled{3}).
This blocking limits inter-stream overlap by preventing the CPU from enqueuing additional work to other streams.
In the timeline for this variant (\autoref{fig:beemon_shuffle}, third plot), this behavior appears as bursts of NIC activity separated by idle periods, reflecting the reduced effectiveness of overlap.

\subsection{Optimization: Deferred Synchronization} 
\label{sub:optimization_deferred_synchronization}
\rev{In TCR-based processing, stream-synchronizing operations are unavoidable for data-dependent operations across streams, making
achieving overlap a unique challenge in this setting.
However, we can reduce the synchronization overhead and achieve better overlap by accounting for relevant operations during scheduling.}
Our key idea is to front-load operations with implicit stream synchronization at the beginning of the work scheduled per stream.
For example, instead of scheduling partitioning, shuffling, and probing for the right-hand chunk \texttt{R1} back-to-back in the same stream, we defer probing \texttt{R1} to avoid forcing an early stream synchronization (see the lower part of \Cref{fig:naive_vs_overlap_figure}).
This strategy ensures that when another stream is used to process the next chunk \texttt{R2}, the asynchronous operations of the previous chunk (e.g., the network shuffle) run concurrently with the synchronizing operations of the current chunk.
In the example, the probe on \texttt{R1} is scheduled as the 10th operation in stream~1, before partitioning and shuffling \texttt{R3}.
We refer to this strategy as the \emph{Deferred Synchronization} variant \circled{4}.
In this variant, hash table operations are deferred: instead of executing them immediately after shuffling within the same stream of the current chunk, they are scheduled as the first operations of the \emph{next} chunk in that stream (see the scheduling order in the lower-right corner of \Cref{fig:naive_vs_overlap_figure}).
This is most beneficial for probing each received chunk of the right-hand table, as described above, but similar measures allow the hash table build to run concurrently with operations on right-hand chunks (e.g., scheduled as the 7th operation after starting the first right-hand shuffle in the figure).

%

\paragraph{Effects of Deferred Synchronization on Overlap.}
Deferred synchronization enables consistent overlap between computation and communication, as visible in the rightmost timeline in \Cref{fig:beemon_shuffle}.
This leads to a substantial runtime reduction and higher effective bus bandwidth, as shown for \emph{Deferred Synchronization} (green bar) in \Cref{fig:shuffle_scalability2}.
Compared to the blocking variant, deferred synchronization improves performance by up to 2.5$\times$.
For larger numbers of columns (i.e., greater shuffle volume) the \emph{Deferred Synchronization} variant approaches the practical upper limit of the network bandwidth, achieving more than 17~GiB/s out of a possible 22.3~GiB/s.

\begin{figure}
    \includegraphics[width=.9\columnwidth]{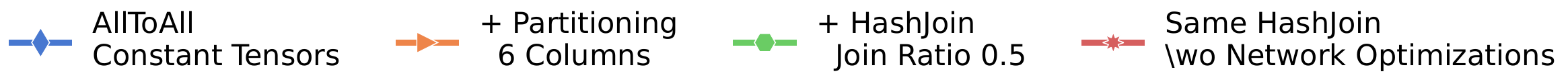}\\\vspace{-0.3em}
    \includegraphics[trim={0 0 0 2em},clip,width=0.8\columnwidth]{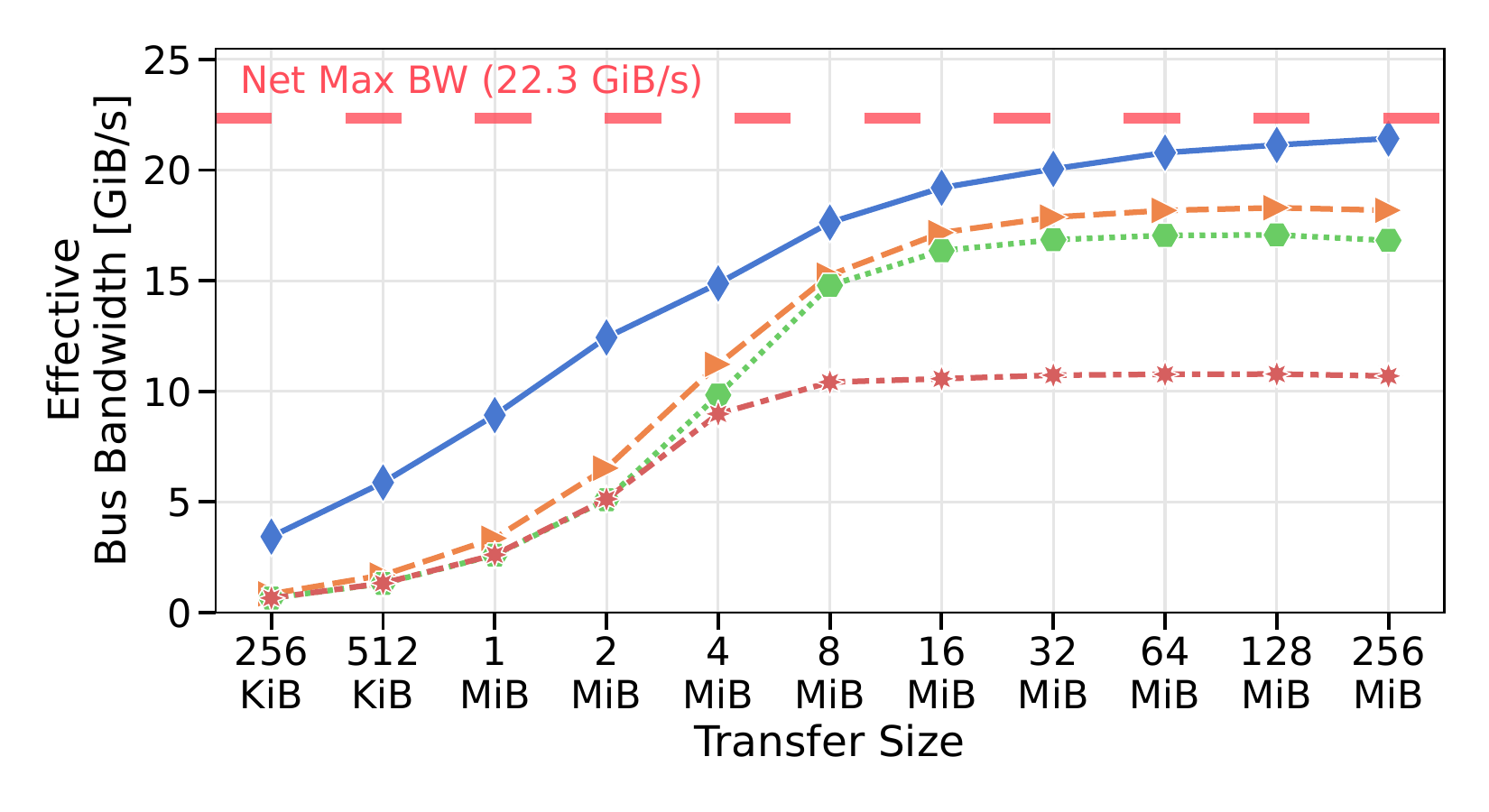}
\vspace{-2ex}
\caption{Impact of transfer size and concurrent computation for a fully overlapped distributed join for 2 Nodes: Bus bandwidth starting from constant-sized transfer without computation, adding concurrent partitioning and join operations. The optimized join (green) is close to the pure shuffle with All-to-All (blue line) = upper baseline.}
\label{fig:shuffle_and_computation}
\vspace{-2ex}
\Description{TODO}
\end{figure}

\paragraph{Remaining Bandwidth Limitations.}
To analyze the remaining performance gap to the theoretical maximum bus bandwidth of the 200~Gbit/s link, we evaluate the overlapped join variants across different transfer sizes.
As shown in \Cref{fig:shuffle_and_computation}, the NCCL \emph{AllToAll} collective alone can reach close to the maximum bus bandwidth for transfer sizes of about 64~MiB (blue line).
However, once the execution stream includes data partitioning, the effective bus bandwidth drops to around 18~GiB/s (orange line) and plateaus for larger transfer sizes.
Running other join phases (build and probe) concurrently with the NCCL kernel further reduces the observed bandwidth slightly (green line).
This reduction stems from the NCCL collective kernel slowing down when it shares GPU resources with other compute kernels (in our case, partitioning), a known form of resource contention when NCCL and computation kernels run concurrently~\cite{nccl_overlap_issue}.
We partially mitigate this effect by increasing the number of thread blocks (NCCL channels) launched as part of our network-parameter optimization, as evidenced by the higher orange and green lines compared to the same join without this change (lower red line).
However, this analysis reveals that the reduced bandwidth is inherent to the concurrent execution of network I/O and computation and cannot be fully mitigated.


\subsection{Benefits of Chunk-Based Execution}

Beyond enabling overlap, our chunk-based execution strategy offers additional benefits over a naive blocking approach.

\begin{figure}
    \includegraphics[width=.9\columnwidth]{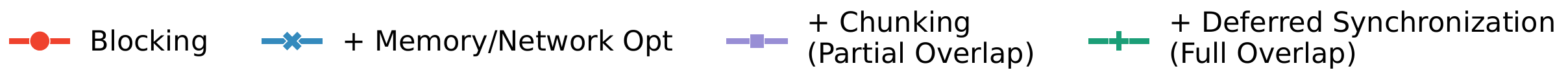}\\
    \includegraphics[trim={2.3em 2.3em 2.3em 2.3em},clip,width=0.48\columnwidth]{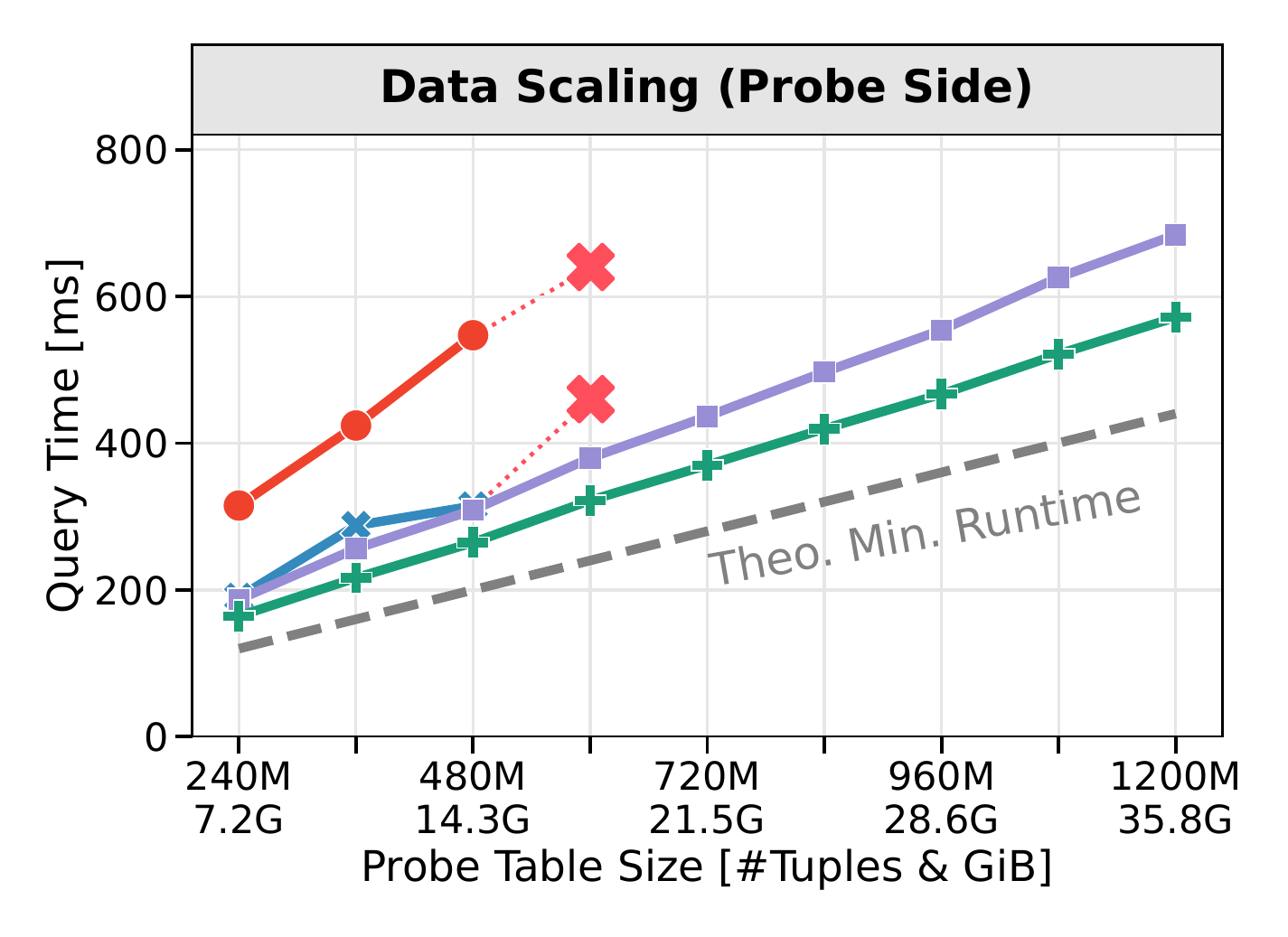}
    \raisebox{0.4em}{
    \includegraphics[trim={2.3em 2.3em 2.3em 2.3em},clip,width=0.459\columnwidth]{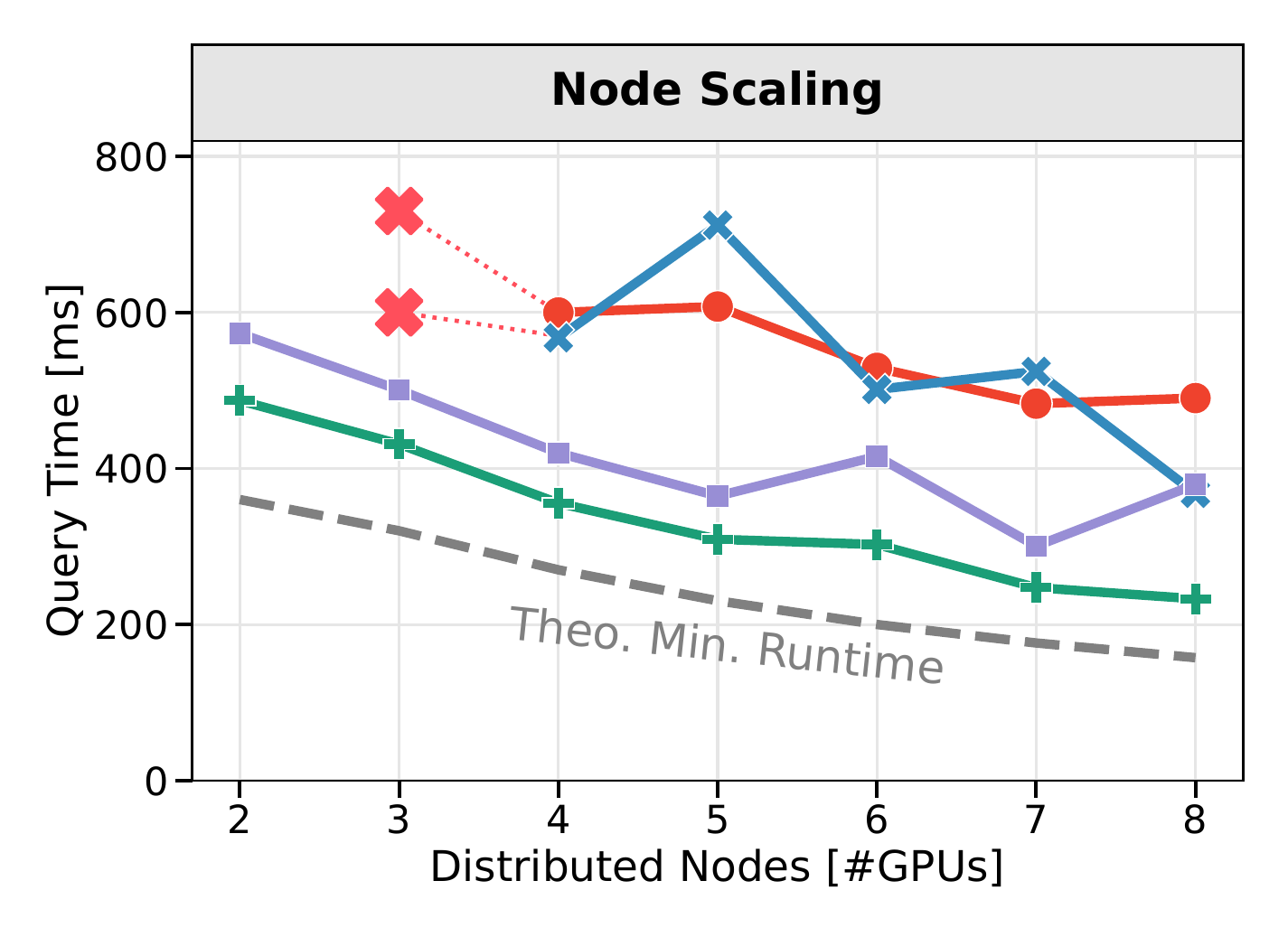}
    }
\caption{\rev{Scaling of a distributed join with four columns and hit ratio 0.5. Left: 2 nodes, 120M build tuples, and increasing probe size. Chunking supports arbitrarily large probe tables as long as intermediates fit in memory. Right: 240M build tuples, 840M probe tuples, and increasing node count. Blocking variants run out of memory at lower node counts and, despite memory and network optimizations, remain slower and more variable than the fully overlapped join.}}
\label{fig:shuffle_scaling}
\vspace{-3ex}
\Description{TODO}
\end{figure}

\paragraph{Reduced GPU Memory Requirements.}
A key advantage of chunk-based execution is its reduced GPU memory footprint during distributed joins.
As long as intermediate results and per-chunk buffers fit into accelerator memory, a chunk-based join can support very large table sizes.
As shown in the left plot of \Cref{fig:shuffle_scaling}, for increasing probe-side table sizes, our chunk-based variants can handle substantially larger probe tables than the blocking baseline due to their pipelined execution model.
Moreover, the overlapped approaches exhibit nearly linear query-time scaling as probe-side data grows, while maintaining a consistently high effective bus bandwidth of approximately 17.2~GiB/s.
For reference, the gray dashed line indicates the theoretical lower bound on runtime, calculated as the per-node shuffle volume divided by the maximum network bandwidth of 22.3~GiB/s.
The remaining gap between the fully overlapped approach and this lower bound is due to the slowdown of collective kernels, as discussed earlier.

\paragraph{Scalability With Increasing Node Counts.}
We also examine scalability with respect to the number of nodes in the distributed setup, keeping the probe- and build-side table sizes fixed.
The right side of \Cref{fig:shuffle_scaling} shows that the overlapped join variants scale effectively as more nodes are added because of their efficient use of collective communication primitives.
The performance of the \emph{Deferred Synchronization} variant approaches the theoretical lower bound (gray dashed line) closely.
Similar to the memory-capacity effect discussed earlier, the blocking join fails to execute at lower node counts because the per-node data volume exceeds the available GPU memory.
For larger node counts, the blocking variants exhibit high runtime variance (e.g., blocking with memory/network optimization, blue line), as slow computation on some nodes leads to stragglers, whereas the fully overlapped variant shows more stable performance.

%% file: section/07_storage.tex

\section{Efficient Storage Access With TCRs} 
\label{sec:storage}

\begin{figure*}[t]
\centering
\includegraphics[width=0.9\textwidth]{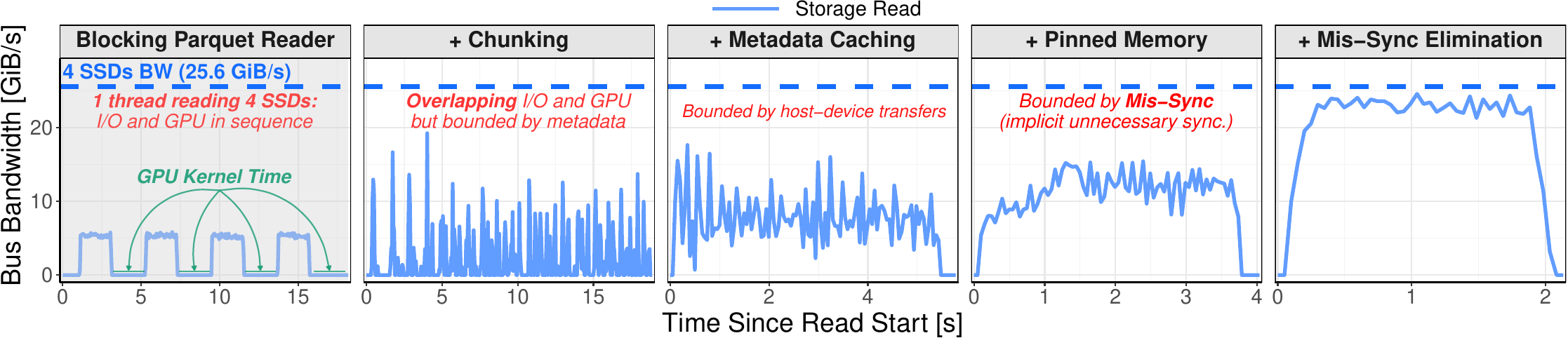}
\caption{GPU Parquet reader scan on TPC-H SF300 \texttt{lineitem}: storage bus bandwidth monitored during reading, with different optimizations. 8 threads are used in all cases except for the blocking reader.}    
\label{fig:storage_bmon}
\end{figure*}

\begin{figure}[t]
 \includegraphics[width=\columnwidth]{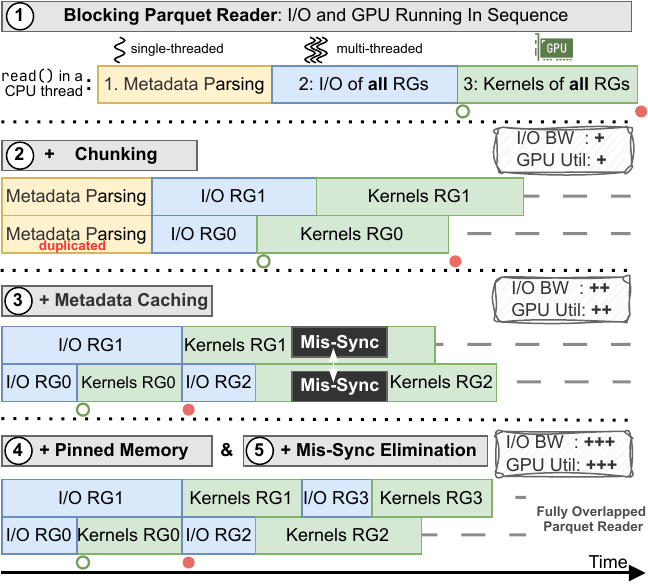}
\caption{GPU Parquet reader: from blocking to fully overlapping storage I/O and GPU with different optimizations}
\label{fig:storage_overlap}
\vspace{-3ex}
\end{figure}

Modern NVMe devices offer high I/O performance at a low cost and are thus also used as primary storage for AI workloads.
However, efficiently exploiting NVMe in TCRs remains under-explored, as TCR research has primarily focused on compute-intensive AI workloads.
In this section, we explore how to effectively integrate NVMe storage into TCRs to implement a table scan operator that reads table files from NVMe and filters tuples using predicates via filter pushdown. 
However, as storage interfaces in TCR are blocking, a table scan first needs to load the entire input into GPU memory before any subsequent operation, 
such as a join, can be applied, resulting in overall non-optimal execution. 
Within the following, we study how an efficient table scan can be implemented on top of PyTorch.
Beyond overlapping I/O and computation, we also study how compression and pruning can further accelerate~scanning.


\subsection{Naive Approach to Table Scans}
\label{subsec:naive_scan}

We now discuss how a first version of a table scan can be realized using PyTorch. 
In PystachIO, we store table in columnar Parquet as a popular open file format.
However, the findings and design decisions we discuss in this section enable efficient table scans with TCRs in general (for other file formats as well).

\paragraph{A Naive Table Scan.} 
A first (naive) approach to implement a table scan is to read table data using an existing Parquet reader.
In our case, we use the GPU-based Parquet reader from RAPIDS cuDF~\cite{libcudf}, 
which supports GPUDirect Storage (GDS)~\cite{gds_website,kvikio_website} for efficient direct I/O and integrates with the PyTorch TCR.
Conceptually, a CPU thread issues the \texttt{read()} in three sequential phases, illustrated as \circled{1} \emph{Blocking Parquet Reader} in~\Cref{fig:storage_overlap}.
In the first phase, a CPU thread parses the Parquet file's metadata~\cite{influxdata2024parquetwide,starburst2025parquetorc,parquet_footer_parsing}.
Second, multiple CPU threads from a pool are used to transfer the requested RowGroups (RGs) to GPU memory for saturating storage bandwidth~\cite{DBLP:journals/pacmmod/BoeschenZB24,DBLP:conf/damon/NicholsonCBA25}.
Finally, GPU kernels decompress and decode the RGs loaded in GPU memory, producing an uncompressed columnar representation for downstream query~processing.
The kernels are launched into the CUDA streams assigned to the CPU thread issuing the \texttt{read()} call.

\begin{figure}[t]
\includegraphics[width=\columnwidth]{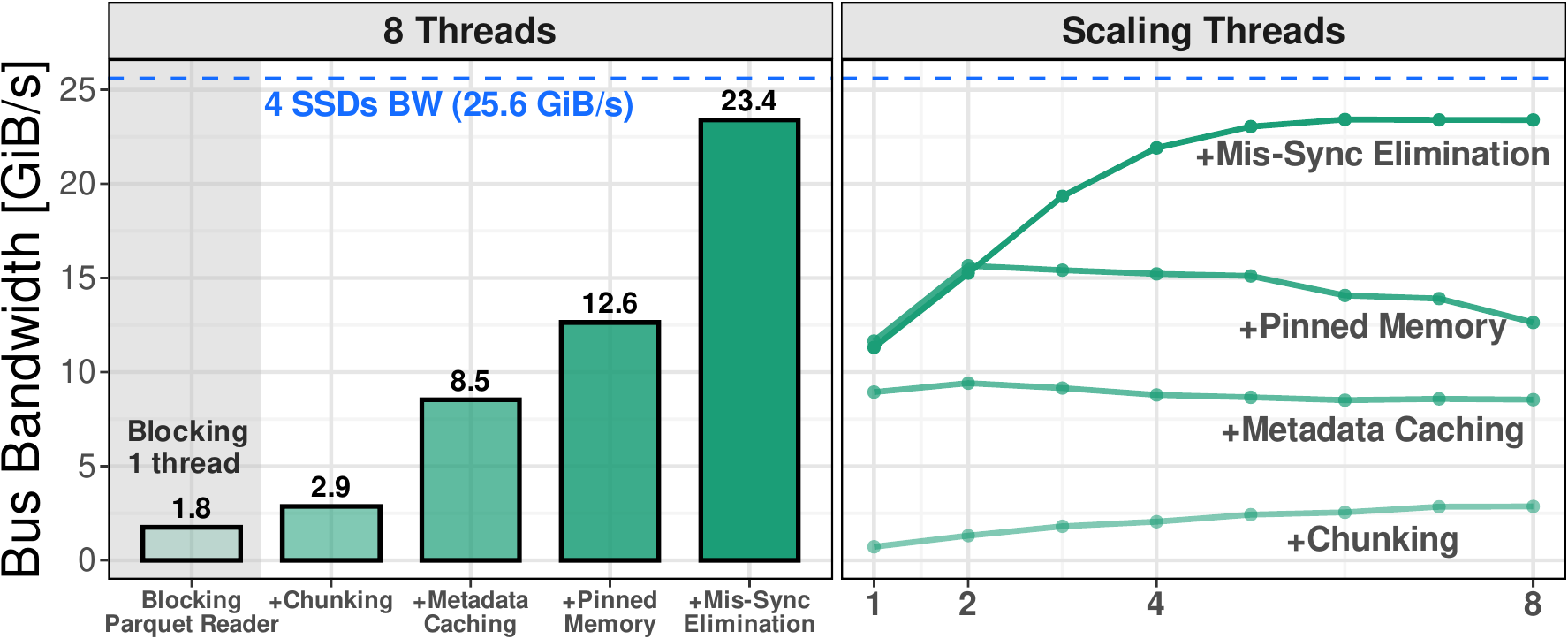}
\caption{GPU Parquet reader scan on TPC-H SF300 \texttt{lineitem}: average storage bus bandwidth under different optimizations. Left: 8 threads are used in all cases except for the blocking reader. Right: Thread scalability from 1 to 8.}
\label{fig:reader_abalation}
\vspace{-5ex}
\end{figure}

\paragraph{Analyzing the Naive Table Scan.}
We identify several limitations of the blocking design used by the current GPU Parquet reader.
First, I/O and GPU execution are strictly serialized: the GPU remains idle during I/O and can launch kernels only afterwards, as indicated by 
\textcolor[HTML]{729D5A}{\raisebox{-0.3ex}{\huge\textbf{\fontfamily{phv}\selectfont $\circ$}}} in~\Cref{fig:storage_overlap}.
As a result, consumer operators (e.g., a join) can begin only after entire columns have been fully materialized in GPU memory, marked with
\textcolor[HTML]{EA6B66}{\raisebox{0ex}{\LARGE\textbf{\fontfamily{phv}\selectfont $\bullet$}}} in~\Cref{fig:storage_overlap}. 
Second, the reader first loads all requested RGs into GPU memory before decompressing and decoding them,  placing substantial pressure on the limited GPU memory and increasing the risk of OOM errors.
Third, the reader can only read one file at a time, limiting the ability to utilize multiple SSDs.
In~\Cref{fig:storage_bmon}, we monitor the storage bus while scanning four files on four separate SSDs using \emph{Blocking Parquet Reader} with one thread.
\rev{Each SSD achieves a peak sequential read bandwidth of 6.4 GiB/s.
Due to the blocking design and the restriction to reading one file at a time, only one SSD is active at any given time during our monitoring.}
Furthermore, during the GPU kernel phase, the storage bus is completely idle in turn.
Overall, when using the \emph{Blocking Parquet Reader} to implement a table scan, 
sequential file reading followed by GPU kernel execution for decompression and decoding, 
the achieved average storage bandwidth is only 1.8~GiB/s, as shown on the left grey side of~\Cref{fig:reader_abalation}.

\subsection{Overlapping Computation and Storage I/O}
\label{subsec:storage_overlap}

To address the limitations of the naive (blocking) table scan for Parquet files, we propose a new Parquet reader for TCRs 
with all optimizations illustrated in~\Cref{fig:storage_overlap}, to fully overlap storage I/O and GPU computation. 
In the following, we first demonstrate that, similar to networking, overlapping is crucial for reducing existing overhead in current GPU query engines with storage capabilities~\cite{velox_cudf, veloxconf_cudf} and afterwards discuss further important optimizations.

\paragraph{Chunking.}

Chunking enables reading column data at finer granularity rather than loading entire columns at once~\cite{DBLP:journals/pacmmod/BoeschenZB24,DBLP:conf/damon/NicholsonCBA25}, 
forming the basis for overlapped I/O and enabling all subsequent optimizations discussed later.
As shown in \circled{2} of~\Cref{fig:storage_overlap}, \emph{Chunking} enables multiple CPU threads to issue \texttt{read()} on different chunks, 
with each thread owning independent CUDA streams for launching kernels.
In addition, chunking also allows parallel reads across multiple SSDs, with CPU threads simultaneously accessing files on different SSDs.
Another important benefit is reduced memory pressure, since chunks are pipelined directly into consumer operators~\cite{DBLP:journals/pvldb/Neumann11} rather than materializing an entire column in GPU memory.
An interesting side effect is that chunking relaxes the row-limit constraints of cuDF, which otherwise restrict the maximum supported read size~\cite{cudf_issue_13159,cccl_issue_47}.  
As illustrated in~\Cref{fig:storage_bmon}, \emph{Chunking}, unlike blocking, enables sustained I/O bandwidth by overlapping computation with data transfer and by reading in parallel from multiple SSDs.  
Although \emph{Chunking} allows short bursts of I/O bandwidth reaching up to 20~GiB/s in~\Cref{fig:storage_bmon}, 
chunking \emph{alone} does not maintain consistently high throughput: the average bandwidth reaches only 2.9~GiB/s, 
as shown in the \emph{Chunking} configuration in~\Cref{fig:reader_abalation}.  
Next, we discuss additional optimizations that enable constant high-throughput I/O in the table~scan.

\paragraph{Metadata Caching.}

Metadata in tabular file formats describes the table schema and provides information that enables data pruning, i.e., identifying prunable RGs from file. 
An inherent overhead of the naive reader is that metadata is re-read in every read request. 
For AI workloads, this overhead is negligible because the read sizes are typically large and not chunked, and the workloads are compute-intensive. 
However, for database workloads, particularly under chunked reads, this results in repeatedly reading and parsing the same metadata for each chunk. 
As illustrated in \circled{2} \emph{Chunking} in~\Cref{fig:storage_overlap}, the same table metadata is redundantly read for every chunk of the same table. 
In PystachIO, we address this inefficiency by caching metadata upfront, thereby decoupling it from the read path. 
Our approach also aligns with common practice in CPU-based Parquet readers~\cite{hao2024caching, datafusionPR16971}. 
Performance-wise, as shown in~\Cref{fig:storage_bmon}, \emph{Metadata Caching} significantly increases sustained I/O bandwidth by decoupling metadata from the critical path. 
With caching enabled, the average read bandwidth reaches 8.5~GiB/s, as shown under \emph{Metadata Caching} on the left of~\Cref{fig:reader_abalation}.

\paragraph{Pinned Memory Usage \& Mis-Sync Elimination.}

After caching metadata, a new performance bottleneck surfaced: unstable multi-stream performance caused by unnecessary synchronizations between CUDA streams. 
These superfluous synchronizations disrupt parallel execution and leave the GPU idle for short periods, as illustrated by the black box in \circled{3} of~\Cref{fig:storage_overlap}. 
This overhead is deeply embedded in the reader, and similar issues have been reported in \texttt{libcudf}-based systems such as RAPIDS accelerated Spark and cuDF-Velox~\cite{velox_cudf,veloxconf_cudf}.

Our analysis shows that these unnecessary synchronizations (mis-sync) are triggered by transfers between pageable CPU memory and GPU memory, for example, when copying small control metadata or element counts typically below a few hundred bytes. 
Although each transfer is tiny, a single such operation in one stream may propagate stalls to other streams~\cite{cudaRuntimeSyncBehavior,cudaProgrammingGuideConcurrentExecution}, 
which is particularly detrimental under our highly parallel chunked execution model.
Because these transfers occur along the control path and involve only small data movements, they are usually overlooked; 
consequently, no prior system has pushed multi-stream reading as aggressively as we do, and this bottleneck has remained unaddressed.
We further identified Thrust~\cite{nvidiaThrust} as the primary source of mis-sync,
producing millions of such synchronization triggers during multi-stream reading.
\rev{To eliminate this bottleneck and achieve full GPU saturation, we \rev{first} enabled CUDA pinned memory and \rev{then} replaced all mis-sync-inducing operations with equivalent implementations using the low-level CUB library~\cite{nvidiaCUB}, thereby avoiding all pageable CPU memory.}
\rev{We follow this order, as pinned memory is a prerequisite for eliminating mis-sync.}
These optimizations yield a streamlined read path with fully overlapped storage I/O and GPU computation, as shown in~\Cref{fig:storage_overlap} (\circled{4} \emph{Pinned Memory} and \circled{5} \emph{Mis-Sync Elimination}), 
resulting in the~\emph{Fully Overlapped Parquet Reader}.

Monitoring the PCIe bandwidth of \emph{Pinned Memory} in~\Cref{fig:storage_bmon} shows that simply enabling pinned memory stabilizes I/O throughput at around 10~GiB/s, 
both significantly higher and more stable than before. 
However, as illustrated on the right side of~\Cref{fig:reader_abalation}, the scalability of \emph{Pinned Memory} alone holds only from one to two threads; 
beyond that, adding threads reduces average read bandwidth. This degradation directly reflects the mis-sync issue: 
with more threads, mis-sync becomes more frequent, severely hurting throughput and even leading to negative scaling.
After applying \emph{Mis-Sync Elimination}, storage bandwidth scales nearly to the full aggregated bandwidth of four SSDs, as shown in~\Cref{fig:storage_bmon} (\emph{Mis-Sync Elimination}). 
Correspondingly, thread scalability in \Cref{fig:reader_abalation} improves dramatically and becomes~near-ideal.


\subsection{Efficient Storage Engine With TCRs}
\label{subsec:storage_scaling}


With all optimizations enabled, our fully overlapped Parquet reader serves as the core of PystachIO's storage engine. 
We believe these contributions advance the design of GPU- and TCR-based storage engines. 
\rev{Several optimizations have already been partially merged into \texttt{libcudf}~\cite{cudfPR18891,cudfIssue18967,cudfIssue18892} and have also motivated broader adoption~\cite{cuCollectionsPR727} among CUDA library developers.}

\begin{figure}[t]
\changedimage{\includegraphics[width=\columnwidth]{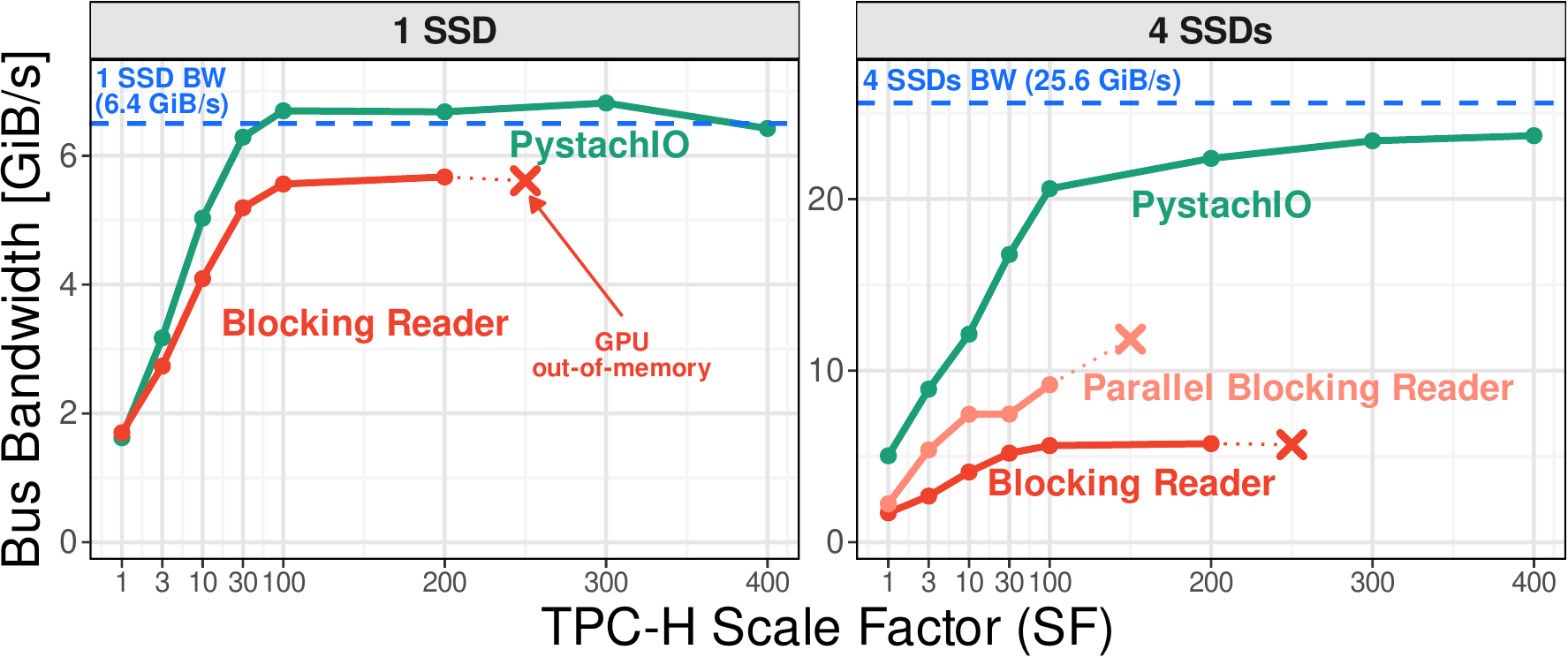}}
\caption{\rev{GPU Parquet reader scan on TPC-H SF300 \texttt{lineitem}: scalability across multiple datasets, compared with the blocking reader and a 4-thread parallel blocking~reader.}}
\label{fig:reader_sf_scaling}
\vspace{-3ex}
\end{figure}

\paragraph{Scalability with Dataset Size.}
\rev{To demonstrate PystachIO as a TCR-backed storage engine, we have evaluated how PystachIO's storage engine scales with increasing TPC-H dataset size.}
As shown in~\Cref{fig:reader_sf_scaling}, PystachIO consistently saturates the available storage bandwidth in both the \emph{1 SSD} and \emph{4 SSDs} configurations. 
At very small scale factors, the input is too small to fully utilize the available bandwidth.
\rev{
In contrast, the blocking reader processes files sequentially, one at a time, and thus achieves substantially lower bandwidth, with limited ability to benefit from multiple SSDs.
Moreover, parallelizing the blocking reader yields only limited further gains. Four parallel threads performing blocking reads across four SSDs can also reach 10 GB/s at SF100, but the parallel blocking reader runs out of memory earlier than the standard blocking reader. This is expected, since in the blocking regime without chunking, higher read throughput also requires substantially larger memory buffers for decompression and~decoding.
}

%% file: section/10_network_and_storage.tex

\section{End-to-end Design With TCRs} 
\label{sec:network_and_storage}

Having demonstrated that TCR-based query processing can efficiently utilize fast networks and storage in isolation, we now investigate how to integrate these insights into an end-to-end TCR query engine that can run distributed queries on SSD-resident data.

\subsection{Naive Integration of Network \& Storage I/O}
\label{subsec:naive_integration}
A naive integration of both storage and network would be to run
Parquet scans from SSDs and distributed processing over RDMA as described in \Cref{sec:storage} and \ref{sec:network} in sequence,
starting networking once a full table is materialized in GPU memory.
To illustrate the impact of this suboptimal approach and further optimizations, we measure the achieved I/O bus bandwidth of different strategies while executing TPC-H query 3 (Q3) as a running example.
We use TPC-H Q3 because it is both storage- and networking-intensive.

\paragraph{Blocking Strategy.}
As a baseline, we execute the query using a single CPU-thread that sequentially performs storage and network I/O, as illustrated in \Cref{fig:storage_and_network_overlap}, top row \circled{1}.
The bandwidth measurement in the top-left plot of \Cref{fig:joint_bmon} shows that this naive strategy does not saturate the bandwidth of fast I/O devices due to a suboptimal interface usage and missing compute and communication overlap, as shown in previous sections.
This blocking strategy yields a query runtime of 6 seconds.

\paragraph{Applying Storage and Network I/O Optimizations.}
To saturate the device bandwidths, we apply the network and storage optimizations introduced in \Cref{sec:network,sec:storage} to the blocking strategy, as visualized by configuration \circled{2} in \Cref{fig:storage_and_network_overlap}.
The top-middle plot in \Cref{fig:joint_bmon} shows the resulting performance.
While the storage bandwidth reaches the device limit, and the query runtime is reduced to 3.3~s, we can observe two remaining challenges:
\textbf{(1)} The storage bandwidth shows oscillations with repeated up-and-down patterns.
\textbf{(2)} The network bandwidth still does not reach the device limit.

To address these challenges, we introduce two further optimizations that are integral to an end-to-end engine to enable efficient overlap of network and storage I/O.

\subsection{Storage Optimization: Reader Combining}
\begin{figure}[t]
\includegraphics[width=.9\columnwidth]{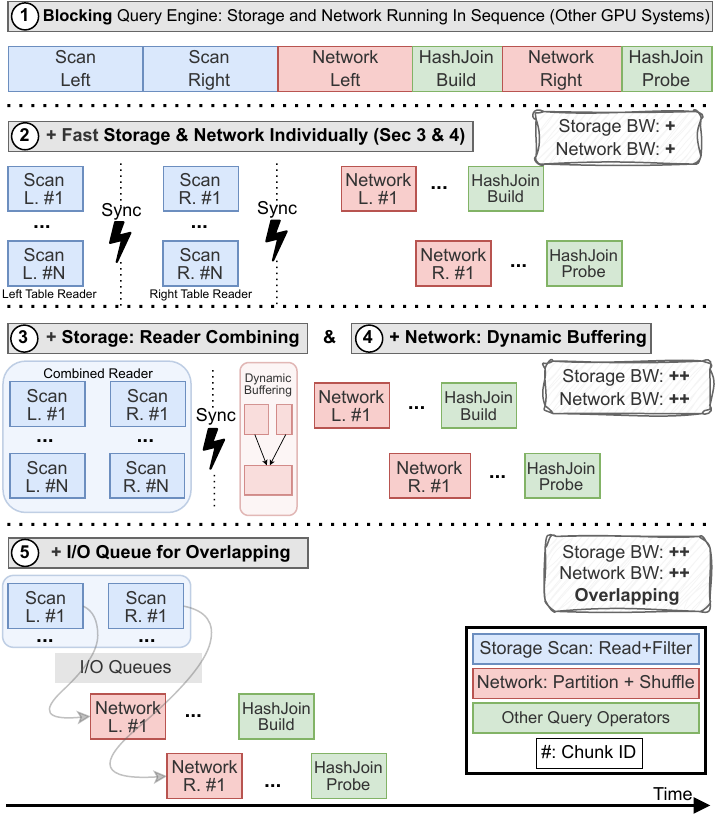}
\caption{Query processing in TCRs: from fully blocking to fully overlapping storage, network I/O, and query execution with different optimizations.}
\label{fig:storage_and_network_overlap}
\Description{TODO}
\vspace{-5ex}
\end{figure}

As previously observed, storage bandwidth oscillates, temporarily underutilizing the device.
These oscillations occur because each table scan operator is executed by a separate Parquet reader with its own storage threads, as described in~\Cref{sec:storage}.
This design serializes table scans: the start and completion of each scan cause corresponding rises and drops in bandwidth.
To address this issue, \emph{reader combining}, illustrated as \circled{3} in \Cref{fig:storage_and_network_overlap}, coordinates table reads through a shared storage I/O thread pool.
Reader combining issues read requests for all table scans in a query plan concurrently, eliminating bandwidth fluctuations and sustaining near-maximum storage bandwidth, as indicated in blue in the top-right plot of \Cref{fig:joint_bmon}. 
A key insight is that global awareness of the entire query plan, rather than optimizing single operators in isolation (e.g., the table scan), is crucial for efficient TCR-based query processing.

\subsection{Network Optimization: Dynamic Buffering}
\label{subsec:dynamic_buffering}
We now address the second inefficiency, the underutilization of network bandwidth.
It arises from transferring data chunks with highly variable and unpredictable sizes.
Chunk sizes vary because network shuffling consumes the output of preceding operators (e.g., filters), whose output size depends on their selectivity.
As shown in \Cref{fig:shuffle_and_computation}, network utilization drops once chunk sizes fall below a few~MiB.
To avoid small chunks and stabilize the transferred chunk size, we introduce a dynamic buffering operator (see \circled{4} in \Cref{fig:storage_and_network_overlap}), denoted as \emph{dynamic buffering}.
We place it before chunk-size-sensitive operators, such as the distributed join, to aggregate incoming chunks until a sufficiently large buffer size is reached.
\Cref{fig:joint_bmon} shows the performance impact of dynamic buffering: the network bandwidth increases from 11.5~GiB/s (top-middle) to 15.5~GiB/s (top-right, in red), thereby saturating the overlapped shuffle bandwidth limit established in \Cref{sec:network}.

\subsection{Overlapping Network and Storage I/O}
\label{sub:overlapping_net_sto}
\begin{figure}[t]
\includegraphics[width=\columnwidth]{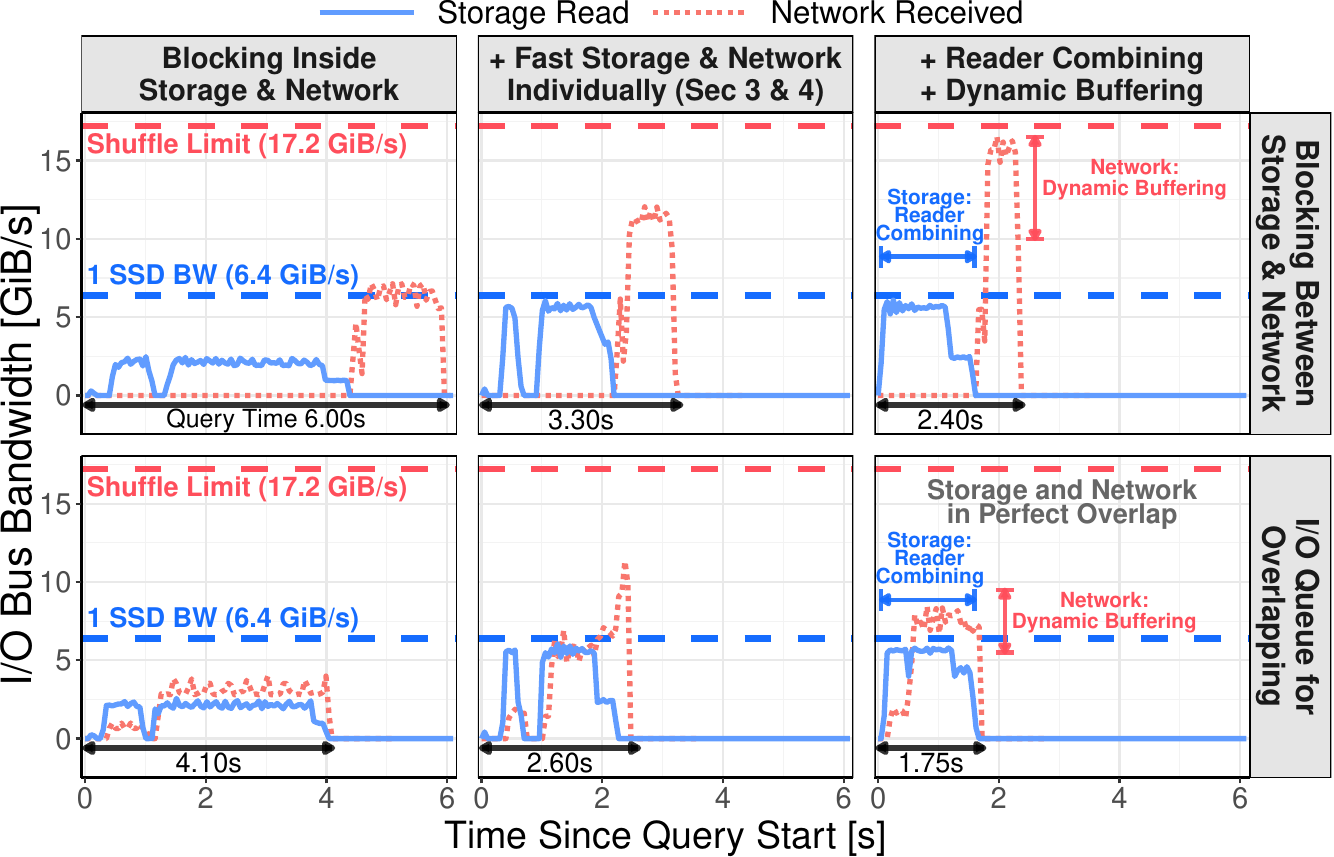}
\caption{TPC-H Query 3 runtime executed on 2 GPUs with I/O performance: storage and network I/O bus bandwidth monitored over the query time with different~optimizations.}
\label{fig:joint_bmon}
\Description{TODO}
\vspace{-6ex}
\end{figure}

The previous sections have shown that integrating network and storage I/O for end-to-end TCR-based query processing requires more than simply executing them sequentially.
Additional optimizations are needed to fully utilize the bandwidth of both subsystems (\Cref{fig:joint_bmon}, top).
PystachIO aims to avoid sequential computation and I/O by overlapping both.
\rev{This allows for reducing memory pressure by discarding fully-processed chunks, keeping only necessary intermediate results.}

\paragraph{Main Challenges of Overlap.}
We identify two main challenges when overlapping network and storage I/O with TCRs.
First, TCR distributed collective operations require execution in the same order across all nodes.
If collective primitives are initiated from different threads, this would require additional synchronization, degrading performance and undermining the simplicity of TCRs.
Therefore, approaches such as a global I/O thread pool shared by network and storage are not feasible.
Second, network and storage I/O have different control-path characteristics.
While network shuffling can be driven efficiently by a single thread, storage I/O for a Parquet table scan typically requires multiple threads to saturate SSD bandwidth (see \Cref{sec:network,sec:storage}).
Because the table scan typically serves as the base operator in query execution, careful coordination is necessary to ensure sufficient resources (e.g., memory) are available for downstream tasks, such as data shuffling.
PystachIO employs two techniques to address these challenges: \emph{queue-based scheduling} and \emph{adaptive regulation}.

\paragraph{Queue-Based Scheduling.}
To address the first challenge and avoid unnecessary thread synchronization, we keep the network and storage I/O paths separate and connect them via a shared multi-threaded queue, as illustrated by \circled{5} in \Cref{fig:storage_and_network_overlap}. 
Such a queue is not provided as a TCR primitive because standard TCR blocking I/O is executed in a single thread and requires no coordination.
We therefore implement a new multi-producer, single-consumer queue as a PyTorch component in PystachIO.
This queue allows multiple CPU threads in the Parquet reader to enqueue filtered data chunks into GPU memory,
while a dedicated networking thread for data shuffling dequeues the chunks for distributed query processing.
With this design, network and storage I/O can progress independently, allowing for full overlap between the two.

\paragraph{Performance Impact of Overlap.}
To quantify the impact of overlap on query execution performance, we analyze I/O-bus bandwidth in the prior experiments with overlap enabled, as shown in the bottom row of \Cref{fig:joint_bmon}.
Enabling overlap with unoptimized I/O (\Cref{fig:joint_bmon}, bottom-left) reduces the query runtime from 6~s to 4~s, even though I/O bandwidth remains underutilized.
Adding the individual I/O optimizations (\Cref{fig:joint_bmon}, bottom-middle) increases storage bandwidth utilization but still suffers from the issues that motivate our reader combining and dynamic buffering techniques.
Combining queue-based scheduling with these optimizations (\Cref{fig:joint_bmon}, bottom-right) reduces the query runtime to 1.75~s.
In this variant, network bandwidth is limited by the preceding storage I/O and therefore stabilizes at 8~GiB/s, which is sufficient to almost completely hide the networking phase.
For the queue variants, data received from storage and network shares a PCIe connection to the GPUs.
However, this potential bottleneck is mitigated since data is loaded in compressed form and decompressed only on the GPU, and because filter pushdown ensures that shuffled data is typically smaller than data ingested from storage.

\paragraph{Adaptive Regulation.}
Beyond decoupling control paths and overlapping I/O, our queue-based scheduling also enables \emph{adaptive regulation} to prevent the second challenge of resource imbalances and GPU out-of-memory (OOM) failures.
OOM can occur in GPU-based query processing when I/O loads data into GPU memory faster than compute operators can consume it.
To mitigate the issue, adaptive regulation associates each I/O queue with a table and assigns a maximum queue size during query planning.
When a queue reaches its limit, storage threads block before producing additional chunks (as a form of backpressure), throttling storage reads if necessary and indirectly aligning the speeds of storage, network, and compute.
PystachIO can robustly process datasets larger than GPU memory without triggering OOM errors in TCRs \rev{with adaptive regulation, incurring negligible storage bandwidth costs, as we will show in the evaluations in the following section}.



%% file: section/60_evaluation.tex


\section{End-to-End Evaluation} \label{sec:evaluation}
To quantify the benefits of PystachIO's optimizations end-to-end, we test its performance and scalability with the TPC-H benchmark.

\begin{figure*}[t]
\centering
\changedimage{\includegraphics[width=\textwidth]{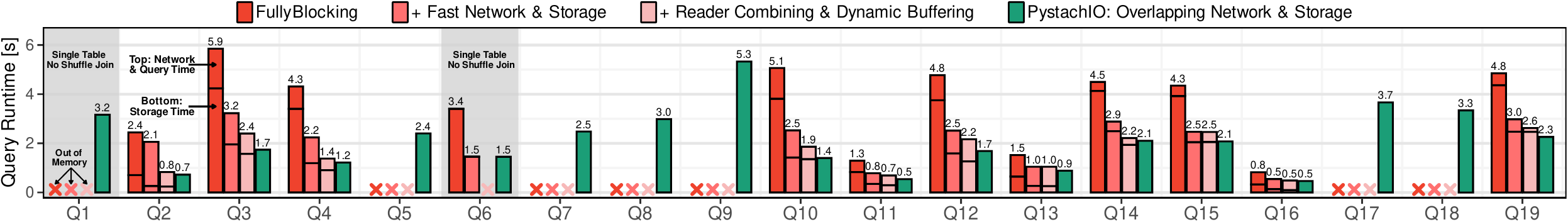}}
\caption{TPC-H SF500 runtime breakdown with 2 GPUs.
\rev{Blocking variants are shown in three shades of red, with each bar decomposed into network and query-processing time (top portion) and storage table scan time (bottom portion).}
}
\label{fig:2gpu-tpch-breakdown}
\vspace{-3ex}
\Description{TODO}
\end{figure*}

\subsection{Experimental Setup}
We first summarize the common experimental settings used in the subsequent experiments.

\paragraph{Hardware.}
All experiments are run on an NVIDIA DGX system~\cite{nvidiaDGXA100Guide} with Ubuntu 22.04 (DGX OS 6), Linux kernel 5.15, CUDA 12.9, and PyTorch 2.10.
The system has two AMD EPYC 7742 64-core CPUs, 8 NVIDIA A100 GPUs with 40~GiB memory each, 8 Mellanox ConnectX-6 200G InfiniBand/Ethernet NICs, and 8 NVMe SSDs with up to 6.4~GiB/s read bandwidth.
Each GPU is paired with its nearest NIC and SSD for network and storage I/O.
Unless noted otherwise, we use 8 CPU threads per SSD for Parquet scans with filter pushdown.
To emulate a distributed environment, we disable NVLink and GPU shared-memory transports, forcing all inter-GPU communication over 200G Ethernet (RoCE) via the NICs.

\paragraph{Workload.}
We use TPC-H to evaluate end-to-end distributed query execution over SSD-resident data.
\rev{We implement all TPC-H queries except Q20, Q21, and Q22, which contain query patterns not yet supported by PystachIO.
Unless noted otherwise, each query uses a distributed symmetric repartitioning hash join for the join between its two largest tables.
These tables are stored as horizontally partitioned Parquet files across the SSDs of different GPUs, while the remaining non-shuffled tables are replicated.
All data is read from SSDs via GDS using Parquet scans with cached metadata.
For queries with string attributes, we dictionary-encode strings as integers and adapt the queries accordingly.
All Parquet files are configured for optimized GPU reading, as described in~\cite{DBLP:journals/corr/abs-2602-17335}}.

\paragraph{Baselines.}
We compare PystachIO’s optimized overlapping approach against three blocking variants with increasing levels of optimization:
(1) \emph{Fully Blocking}: Single-threaded execution for both storage and networking.  
(2) \emph{Fast Network \& Storage}: All network and storage optimizations from \Cref{sec:network,sec:storage}. 
(3) \emph{Reader Combining \& Dynamic Buffering}: Additionally enables reader combining and dynamic buffering from \Cref{sec:network_and_storage}.

\rev{As a reference, we estimate that the runtime of a system combining the distributed TCR query execution described by \citeauthor{DBLP:journals/corr/abs-2506-09226} (column-granular blocking execution) with a previous highly efficient storage phase 
would result in a performance close to baselines (2) and (3).}
\rev{In the final experiment, we compare PystachIO against the performance of the CPU systems DuckDB and ClickHouse.}

\paragraph{Performance Upper Bound.}
As an upper bound on performance, we compare PystachIO against a theoretical minimum query runtime, $T_{min}$, derived from the data size and hardware bandwidth:\vspace{-1ex}
\[T_{min} = \max\left(\frac{\text{SSDReadSize}_{agg}}{\text{SSDReadBW}_{agg}}, \frac{\text{NetRecvSize}_{node}}{\text{NetBW}}\right)\tag{1}\label{eq:tmin}\vspace{-1ex}\]
The first term of \Cref{eq:tmin} bounds the runtime by the time to read the total compressed input $\text{SSDReadSize}_{agg}$ from storage at full aggregated SSD bandwidth $\text{SSDReadBW}_{agg}$.
For example, with 12.8~GiB of compressed data and two SSDs providing a total read bandwidth of 12.8~GiB/s (6.4~GiB/s each), the minimum runtime is 1 second.
This bound corresponds to a system that processes queries at data-loading speed, with no decompression, network, or query-processing overhead.
As we scale to more nodes (i.e., more GPUs in our setup), query runtime can become network-bound and dependent on how much data $\text{NetRecvSize}_{node}$ a node can receive at network speed $\text{NetBW}$. This lower query runtime does not account for contention on the GPUs' PCIe links, making it an even stronger lower bound.
We report $T_{min}$ as \emph{Theoretical Min. Runtime} in our experimental results.

\subsection{\rev{Query Runtime Breakdown}}
\rev{We begin by evaluating PystachIO's performance on all supported TPC-H queries to provide an overview using 2 GPUs on SF500 (corresponding to a 500 GB dataset).
To analyze the performance gains from overlapping storage and network I/O, \Cref{fig:2gpu-tpch-breakdown} compares the fully overlapped against the runtime breakdown of the three blocking variants}.
We observe that:
\textbf{(1)} Optimizing storage and network I/O in isolation already reduces query runtimes substantially (configuration 2).
\textbf{(2)} Storage and network \& query time often account for comparable fractions of the total runtime, indicating that overlapping these phases offers substantial gains.
\textbf{(3)} PystachIO's query runtime is only slightly higher than the storage time of the fastest blocking variant (configuration 3) and is close to the maximum of the storage and network times, implying close overlap of the two phases (i.e., the shorter phase is effectively hidden).
\rev{\textbf{(4)} Beyond runtime benefits, our overlapping approach also helps avoid OOM errors, as we explain in the next~subsection.}
\paragraph{Takeaway:} Overlapping storage and network I/O largely hides the shorter phase and significantly improves performance.

\subsection{\rev{Dataset Scalability}}
\begin{figure}
\centering
\changedimage{\includegraphics[width=\columnwidth]{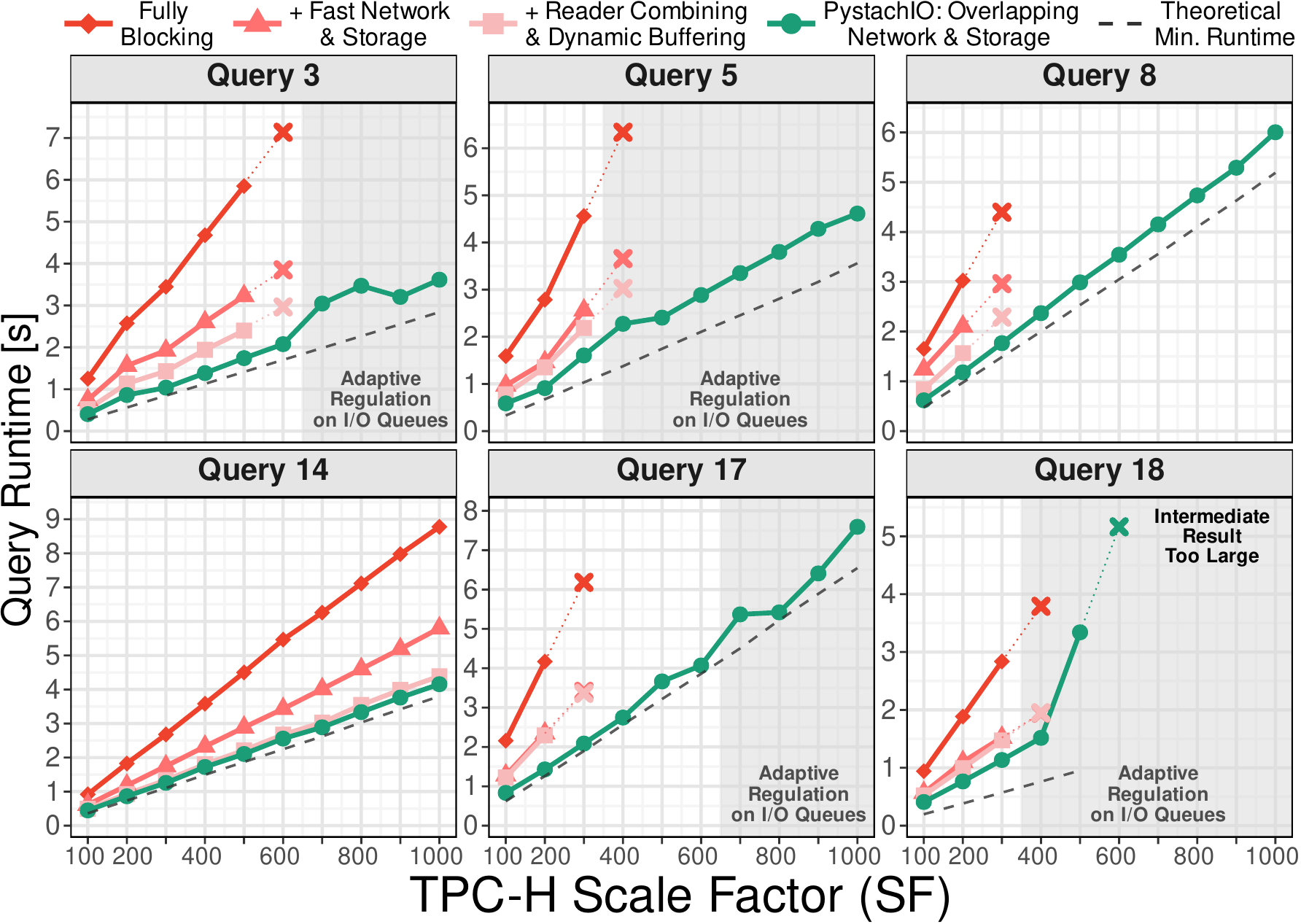}}
\caption{Dataset scalability up to TPC-H SF1000: query performance with~2 GPUs.}
\label{fig:2gpu-tpch}
\Description{TODO}
\end{figure}

\rev{Next, we study PystachIO's performance behavior as the data size increases.
\Cref{fig:2gpu-tpch} reports the end-to-end runtimes of six representative TPC-H queries} for scale factors from 100 to 1000 (corresponding to 100~GB to 1~TB).
We summarize our main observations as follows:

\textbf{(1)} PystachIO with all optimizations (green line) scales linearly with the SF and is close to the theoretical minimum.
\textbf{(2)} The blocking variants run out of GPU memory as the dataset size increases \rev{(Q3, Q5, Q8, Q17)}, because their table scans attempt to materialize the full input in GPU memory.
\rev{\textbf{(3)} For highly-selective queries with little shuffled data, such as \rev{Q14}, the workload is primarily storage-bound and the networking and query processing phase is short,
leading to both PystachIO and the optimized blocking variant being able to approach the minimum.}
\textbf{(4)} For queries with heavier shuffling, PystachIO benefits from our adaptive regulation technique (highlighted with a gray background), which limits the I/O queue size to avoid GPU OOM at the cost of slightly slower storage throughput \rev{(see lightly-widening gap between PystachIO and optimum, e.g., for Q3 and Q5)}.
\rev{\textbf{(5)} Opportunities for further optimizations become apparent for Q18, which suffers from operator overheads in the aggregation and large intermediate results. These situations motivate further research into more processing- and memory-efficient TCR-based query operators.}

\paragraph{Takeaway:} PystachIO is highly efficient and close to the theoretical minimum. Our overlapping technique improves runtime and reduces memory consumption, mitigating the OOM problems in large-scale data processing.

\subsection{Cluster Scalability}
We evaluate how PystachIO scales with the number of processing nodes using up to all 8 GPUs on the DGX system. 
For the dataset size, we use the largest scale factor, SF1000, to determine at which cluster size the blocking variants overcome OOM failures.
The results in \Cref{fig:scaling-gpu-tpch} show that:
\textbf{(1)} Overlapping storage and network I/O, along with adaptive queue regulation, enables the execution of large datasets with significantly fewer nodes.
For example, the blocking variants require at least 8 nodes to fully materialize the inputs of \rev{Q5, Q8, and Q17, whereas PystachIO processes these queries efficiently with only 2 nodes by using adaptive regulation of the I/O queues (which is not possible with the blocking variants).
Only Q18 requires at least 6 GPUs for execution, due to the operator and memory overheads discussed before, leading to gaps in the optimal runtime.
}
\textbf{(2)} PystachIO scales efficiently with an increasing number of GPU nodes, as the underlying storage and network collective communication primitives are highly scalable, and PystachIO can effectively orchestrate and overlap them.
\textbf{(3)} Fully overlapped execution remains close to the theoretical minimum runtime as the number of nodes increases.
\rev{Only Q18 shows the performance gap of large materialized aggregation, but the scaling to 8 nodes exhibits the expected decrease in runtime and decreases the size of the gap.}

\paragraph{Takeaway:} For TCR-based distributed query processing, PystachIO scales efficiently with node number.
By leveraging fast storage and overlapping I/O, it can process large datasets with a small number of nodes, enabling cost-efficient large-scale query~processing.

\begin{figure}
    \centering
    \changedimage{\includegraphics[width=\columnwidth]{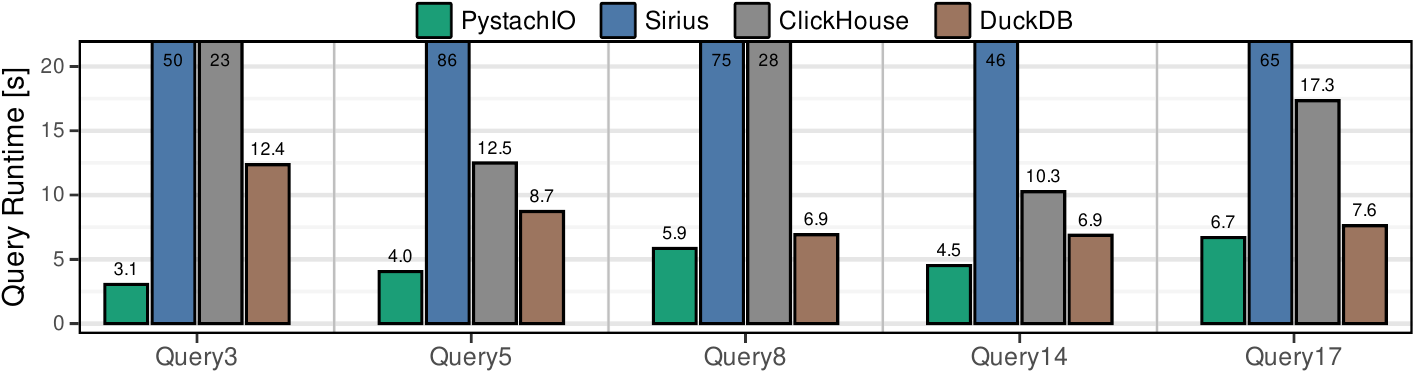}}
    \\[.5em]
    \changedimage{\includegraphics[width=\columnwidth]{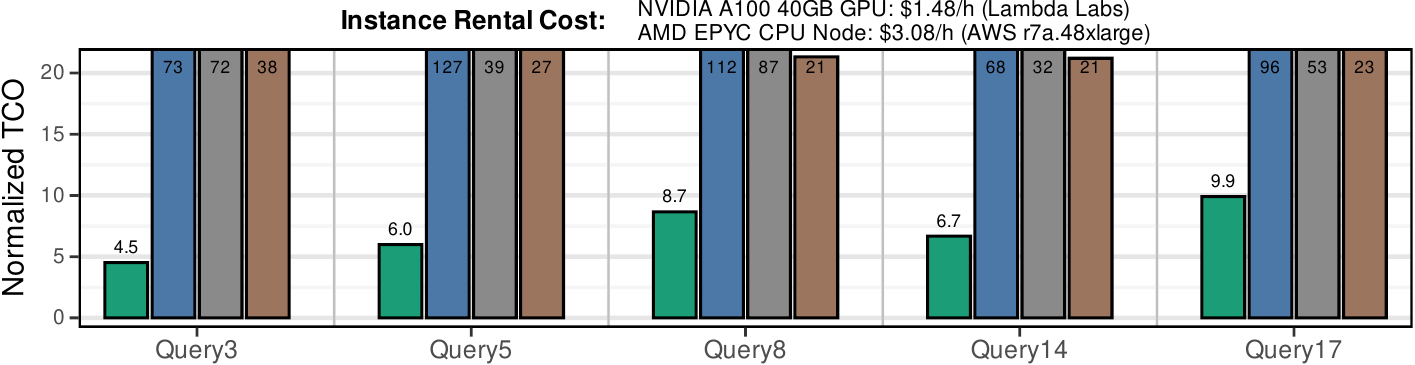}}
    \caption{System comparison on TPC-H SF500 runtime and normalized TCO in a single-node setting with fully storage-resident data.}
    \label{fig:tpch_breakdown_runtime_tco_cold}
    \Description{TODO}
\end{figure}


\begin{revision}
\subsection{Systems Comparison}
\label{sec:eval:system_comparison}
Finally, we compare PystachIO against DuckDB~\cite{DBLP:conf/sigmod/RaasveldtM19} and ClickHouse~\cite{DBLP:journals/pvldb/SchulzeSYDM24}, two CPU-based systems known for strong OLAP performance, including on storage-resident workloads.
We also compare against Sirius~\cite{DBLP:journals/corr/abs-2508-04701}, a GPU system that relies on CPU-based DuckDB for Parquet ingestion. 
We restrict this comparison to a single node because, unlike PystachIO, these systems do not support a distributed repartitioning join. 
As a result, this experiment isolates single-node query processing over SSD-resident data rather than distributed execution.

The runtimes in the upper part of \Cref{fig:tpch_breakdown_runtime_tco_cold} show that PystachIO remains highly competitive even in this setting. 
DuckDB achieves similarly low runtimes for several queries, reflecting its efficient storage layer. 
However, PystachIO outperforms the other systems on queries with large scan inputs, such as Q3 and Q5, where its overlap of storage access and GPU execution is particularly effective. 
Sirius is also notable: although it often matches or exceeds PystachIO on in-memory workloads, it performs worse here because data loading and preparation remain CPU-driven, introducing substantial overhead for storage-resident execution.

The lower part of \Cref{fig:tpch_breakdown_runtime_tco_cold} converts these runtimes into total-cost-of-ownership (TCO) estimates using cloud VM prices for hardware comparable to our setup, following the methodology of~\cite{DBLP:journals/corr/abs-2508-04701}. 
Because PystachIO combines low query runtimes with competitive pricing for GPU-equipped instances, it achieves at least 2.5$\times$ lower TCO on the evaluated queries.

\begin{figure}
\centering
\changedimage{\includegraphics[width=\columnwidth]{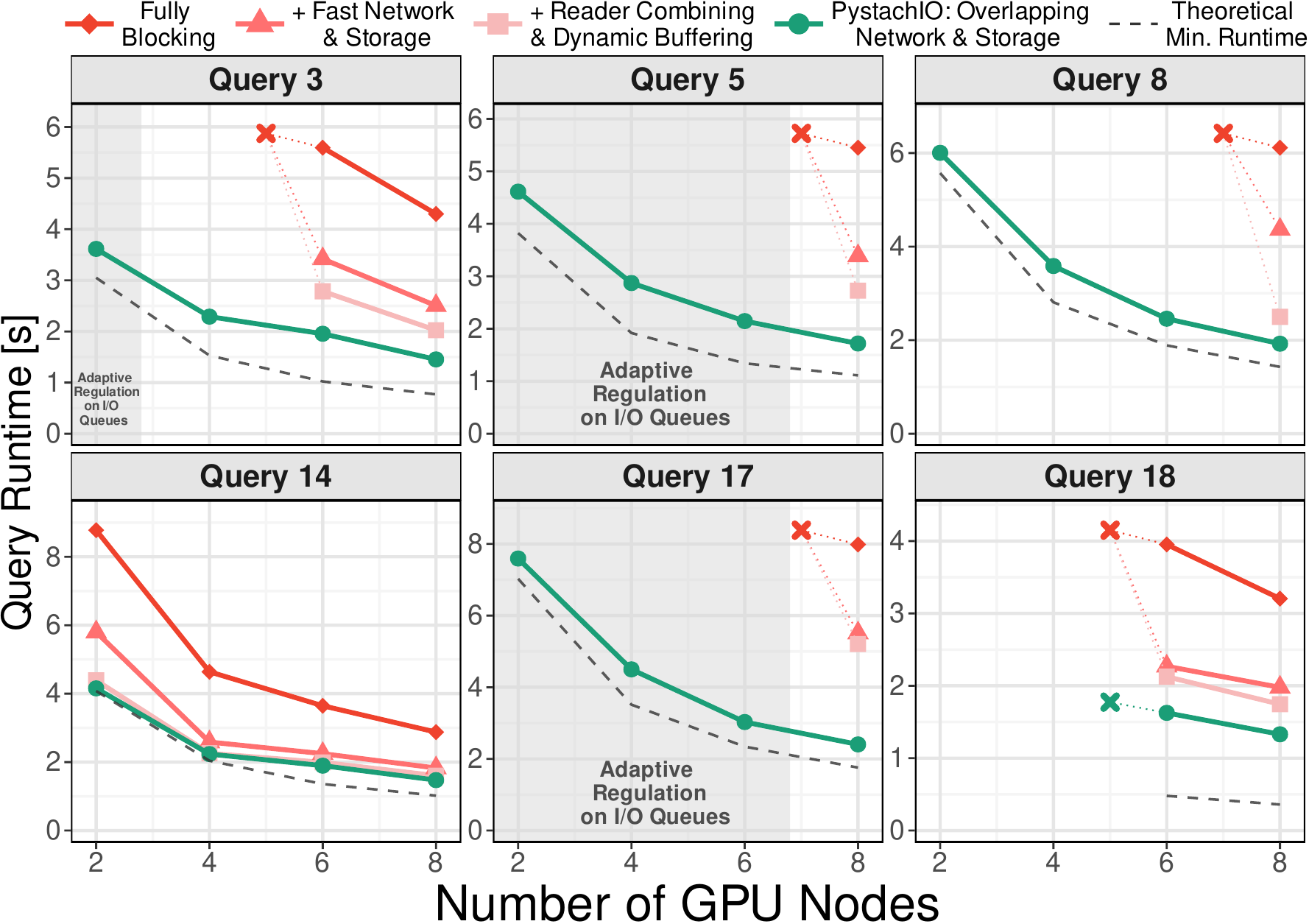}}
\caption{GPU-node scalability up to 8 GPUs: query performance with TPC-H SF1000.}
\label{fig:scaling-gpu-tpch}
\Description{TODO}
\vspace{-3ex}
\end{figure}

\end{revision}

%% file: section/70_related.tex

\section{Related Work} \label{sec:related}

\paragraph{GPU Query Processing}
Many works~\cite{DBLP:conf/sigmod/LutzBZRM20,DBLP:conf/sigmod/LutzBZRM22,DBLP:conf/sigmod/BressFT16,DBLP:conf/sigmod/ShanbhagMY20,DBLP:journals/pacmmod/KabicCA25} have proven the acceleration potential of GPUs for analytical query processing. 
Among them, the TQP series~\cite{DBLP:journals/pvldb/HeNBSSPCCKI22,DBLP:journals/pvldb/CaoSIAK23,DBLP:journals/corr/abs-2506-10092,DBLP:journals/pvldb/LiDWMABBCIPRCG25} pioneered TCRs as execution engines and optimized their performance for in-memory scenarios.
We use the TCR-based execution model for our system, but adapt it for
chunked execution and target
I/O-heavy queries over fast networks and fast storage.

\paragraph{Extending GPU Query Processing Over Networks}
Distributed GPU query processing has been studied in several systems~\cite{DBLP:conf/adms/GaoS21,DBLP:conf/icpp/GuoCZL19,DBLP:journals/pacmmod/ThostrupDBLB23,DBLP:journals/corr/abs-2508-04701} that focus on GPU-resident or hybrid CPU--GPU memory workloads, do not use TCRs for execution, and do not consider storage.
Theseus~\cite{DBLP:journals/corr/abs-2508-05029} explores distributed execution together with data loading from SSDs; however, its system architecture has not yet been described in detail, which limits a thorough comparison with our approach.
\citeauthor{DBLP:journals/corr/abs-2506-09226}~\cite{DBLP:journals/corr/abs-2506-09226} deploy a TCR-based database on multi-node GPU clusters with NVLink and RDMA for communication, and demonstrate large-scale query execution.
The key differences are that they (i) combine intra-node interconnects with inter-node communication, (ii) minimize network traffic by co-partitioning tables, and (iii) assume that table data is materialized in the aggregate GPU memory.
By contrast, we focus on distribution over RDMA networks, data loading from storage, and overlapping I/O with computation in TCRs.
We expect their approach to approximate the networking performance of our blocking variants, since they process and shuffle data at column granularity.

\paragraph{Extending GPU Query Processing to Storage}
While GPU networking has been well studied, relatively few works have examined storage integration in GPU query processing, and none have considered TCRs. 
\cite{DBLP:journals/pacmmod/BoeschenZB24,DBLP:conf/damon/NicholsonCBA25} and \cite{DBLP:conf/damon/AfroozehFB24,DBLP:conf/damon/HepkemaAFBM25} study compression techniques and encodings for accelerated analytical processing, respectively.
Although these approaches demonstrate the benefits of GPU-based decompression for specialized formats, industry systems~\cite{nvidiaRapidsSpark,velox_cudf,veloxconf_cudf,nvidia_rapids_dask,polars_gpu_release,DBLP:journals/corr/abs-2508-05029} increasingly focus on end-to-end query performance when loading data from open file formats such as Parquet.
To our knowledge, we are the first to study fast storage access explicitly designed to overlap I/O with query processing in TCRs.

%% file: section/80_conclusion.tex

\section{Conclusion} \label{sec:conclusion}
We presented PystachIO, a \rev{prototypical} GPU-accelerated distributed analytical query engine built on the PyTorch TCR that exploits fast RDMA-capable networks and NVMe storage.
By analyzing why naive use of TCR I/O abstractions underutilizes modern hardware, we derived optimizations that overlap storage and network I/O with GPU computation, achieving high throughput and low end-to-end query runtimes on large datasets. 